\documentclass{article}
\usepackage{moreverb}
\usepackage{array}
\usepackage{tikz}
\usepackage[numbers]{natbib}
\usepackage{booktabs}
\usepackage{amsmath}
\usepackage{amsfonts}
\usepackage{algorithm}
\usetikzlibrary{arrows.meta}
\usepackage{graphicx} %% for Jpegs\usetikzlibrary{arrows.meta}s
\usetikzlibrary{shapes,positioning}
 \usepackage{url}
\usepackage{moreverb}
\pdfoutput=1 
\newcommand\BibTeX{{\rmfamily B\kern-.05em \textsc{i\kern-.025em b}\kern-.08em
T\kern-.1667em\lower.7ex\hbox{E}\kern-.125emX}}

\usepackage{subcaption}

\usepackage{enumitem}

\usepackage{algpseudocode,algorithm}

\makeatletter
\algrenewcommand\ALG@beginalgorithmic{\normalfont}
\makeatother

\newcommand{\pkg}[1]{{\normalfont\fontseries{b}\selectfont #1}}  

\setlist[description]{font=\normalfont\itshape\textbullet\space}

\newcommand\circledcheck[1][cyan!50]{%
  \tikz\node[circle,fill=#1,inner sep=0.8pt]{${\checkmark}$};%
}

\newcommand\circledO[1][green!50]{%
  \tikz\node[circle,fill=#1,inner sep=4.3pt]{\textbf{}};%
}

\newcommand\circledX[1][red!50]{%
  \tikz\node[circle,fill=#1,inner sep=0.8pt]{$\times$};%
}

\begin{document}

\title{Using Individualized Treatment Effects to\\ Assess Treatment Effect Heterogeneity}

\date{}
\author{Konstantinos Sechidis\thanks{Advanced Methodology and Data Science, Novartis Pharma AG, Basel, Switzerland}
\and Cong Zhang\thanks{China Novartis Institutes for Bio-medical Research CO., Shanghai, China}
\and Sophie Sun\thanks{Advanced Methodology and Data Science, Novartis Pharmaceuticals Corporation, East Hanover, New Jersey, USA}
\and Yao Chen\footnotemark[3]
\and Asher Spector\thanks{Department of Statistics, Stanford University,
Stanford, California, USA}
\and Björn Bornkamp\footnotemark[1]
}

% \presentaddress{This is sample for present address text this is sample for present address text.}

%\fundingInfo{Text}
%\JELinfo{ejlje}

\maketitle
\abstract{Assessing treatment effect heterogeneity (TEH) in clinical trials is crucial, as it provides insights into the variability of treatment responses among patients, influencing important decisions related to drug development. Furthermore, it can lead to personalized medicine by tailoring treatments to individual patient characteristics. This paper introduces novel methodologies for assessing treatment effects using the individual treatment effect as a basis. To estimate this effect, we use a Double Robust (DR) learner to infer a pseudo-outcome that reflects the causal contrast. This pseudo-outcome is then used to perform three objectives: (1) a global test for heterogeneity, (2) ranking covariates based on their influence on effect modification, and (3) providing estimates of the individualized treatment effect. We compare our DR-learner with various alternatives and competing methods in a simulation study, and also use it to assess heterogeneity in a pooled analysis of five Phase III trials in psoriatic arthritis. By integrating these methods with the recently proposed WATCH workflow (Workflow to Assess Treatment Effect Heterogeneity in Drug Development for Clinical Trial Sponsors), we provide a robust framework for analyzing TEH, offering insights that enable more informed decision-making in this challenging area.}


% \jnlcitation{\cname{%
%\author{Taylor M.},
%\author{Lauritzen P},
%\author{Erath C}, and
%\author{Mittal R}}.
%\ctitle{On simplifying ‘incremental remap’-based transport schemes.} %\cjournal{\it J Comput Phys.} \cvol{2021;00(00):1--18}.}

\maketitle

\renewcommand\thefootnote{}

\renewcommand\thefootnote{\fnsymbol{footnote}}
\setcounter{footnote}{1}

\section{Introduction}
\label{sec:intro}
The assessment of treatment effect heterogeneity (TEH) is crucial for advancing drug development and personalized healthcare.\cite{rube:shen:2015} TEH facilitates subgroup identification based on baseline covariates, allowing scientists to pinpoint variations in treatment efficacy and safety, which is invaluable for optimizing clinical trial design and stratifying patient groups.\cite{lipkovich2017tutorial} % This assessment helps understand efficacy results and identify suitable patient populations for treatment.\cite{alosh:2017} 
Additionally, TEH enables healthcare providers to tailor medical interventions to an individual's unique biological makeup, lifestyle, and environmental factors, ensuring the most effective and safest treatment.\cite{Wang2023} This personalized approach represents a shift from the traditional ‘one-size-fits-all’ model to a more nuanced, patient-centric model, promising improved health outcomes and a higher standard of care.\cite{Dahabreh2016}

%The assessment of treatment effect heterogeneity (TEH) is an important element in advancing drug development, and ultimately can lead to personalized healthcare.\cite{rube:shen:2015} In terms of the first, TEH facilitates subgroup identification, i.e. identifying groups of patients, based on the baseline covariates, that experience desirable characteristics, such as an enhanced treatment effect.\cite{lipkovich2017tutorial} By identifying these subgroups, scientists can pinpoint variations in treatment efficacy and safety, which is invaluable for optimizing clinical trial design and stratifying patient groups. This assessment is crucial for understanding the efficacy results and, consequently, identifying the suitable patient population for the treatment.\cite{alosh:2017}

%Furthermore, TEH assessment can help healthcare providers tailor medical interventions to the individual’s unique biological makeup, lifestyle, and environmental factors.\cite{Wang2023} This level of customization ensures that each patient receives the most effective and safest treatment, maximizing therapeutic benefits while minimizing potential risks. Personalized healthcare, rooted in the assessment of heterogeneity, represents a paradigm shift from the traditional ‘one-size-fits-all’ approach to a more nuanced and patient-centric model, promising improved health outcomes and a higher standard of care for all.\cite{Dahabreh2016}

Assessing TEH using clinical trial data is challenging due to several factors. Firstly, sample size limitations in clinical trials often prevent definitive statements about testing interactions or estimating treatment effects in subgroups, as the trials are typically designed to demonstrate effects in the overall population.\cite[Chapter 16]{gelman2020regression} Secondly, multiplicity issues arise when assessing treatment effects across various subgroups, leading to selection bias, and findings that cannot be replicated in subsequent studies.\cite{bretz2014multiplicity} Despite these challenges, understanding and accurately interpreting the observed TEH is crucial, as it can influence drug development decisions and future trial designs.

In an attempt to tackle the main challenges around assessing TEH, Sechidis et al.\cite{sechidis2025watchworkflowassesstreatment} introduced a Workflow for Assessing Treatment effeCt Heterogeneity (WATCH) designed for clinical trial sponsors. The main goal of WATCH is to provide a systematic approach for sponsors to make informed decisions based on treatment effect heterogeneity, considering external evidence and scientific understanding. It consists of four main steps: Analysis Planning, Initial Data Analysis and Analysis Dataset Creation, TEH Exploration, and Multidisciplinary Assessment (Figure \ref{fig:workflow}).  The TEH exploration is the core analytical part of the workflow, and it is focusing on addressing three key objectives: 
\begin{figure}
\centering
\centering
\includegraphics[width=0.45\textwidth]{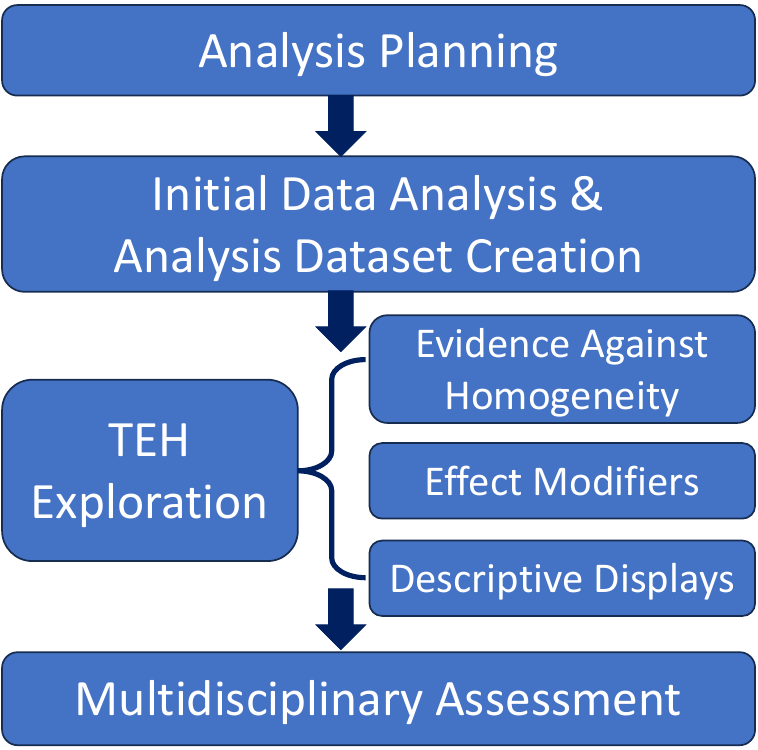}
  \caption{Overview of WATCH workflow and the four main steps: (1) Analysis Planning, (2) Initial Data Analysis and Analysis Dataset Creation, (3) TEH Exploration, and (4) Multidisciplinary Assessment.}
  % \vspace{-2.2cm}
  \label{fig:workflow}
\end{figure}
{ 
\begin{description}
    \item[Objective 1:] perform a global test to assess the evidence against homogeneity. This test serves as a diagnostic tool to summarize the overall signal of effect modification across covariates. However, it is not intended as a gatekeeper for subsequent analyses; downstream steps proceed regardless of the test outcome, ensuring that potentially meaningful modifiers are not overlooked due to limited power.
    \item[Objective 2:] derive a ranking of the baseline covariates that captures their strength in the effect modification. We generate a ranked list without imposing a fixed selection threshold. This approach avoids arbitrary cutoffs and supports a multidisciplinary evaluation that integrates statistical evidence with external data, such as clinical insight and biological plausibility.
    \item[Objective 3:] explore how the treatment effect varies with the most promising effect modifiers. This includes both descriptive approaches, such as per-arm response plots, and model-based strategies that estimate individualized and subgroup-specific treatment effects. The latter requires reliable estimation methods.
\end{description} }

WATCH is a generic workflow, and it can be used with any method that can address the three objectives above. 
In this paper we will explore implementations building upon causal inference, and more specifically on the potential outcomes framework. This framework provides a powerful tool to conduct individual-level analysis to assess TEH and understand the differential impacts of treatments across diverse patient characteristics.\cite{Sanchez2022} Over the last few years, a plethora of methods have been suggested to estimate individualized effects, such as the Conditional Average Treatment Effect (CATE) using various types of models, ranging from traditional regression techniques to more advanced machine learning (ML) algorithms.\cite{Curth2024} Among these advanced approaches, meta-learners have emerged as a powerful framework for estimating CATE, leveraging the strengths of multiple models to improve predictive accuracy and robustness in treatment effect estimation.\cite{kunzel2019metalearners, jacob2021cate, curth2021nonparametric}

In this paper, we will focus on a specific type of meta-learner, known as Doubly Robust (DR) learner,\cite{kennedy2023towards}  which combines the strengths of both outcome modeling and propensity score modeling to achieve more accurate estimates of CATE. DR-learner is particularly advantageous because it remains consistent if either the outcome model or the propensity score model is correctly specified, offering some safeguards against model misspecification. While it was originally introduced to estimate CATE (Objective 3), we will illustrate ways in which it can also be used to perform a global test against homogeneity (Objective 1) and to identify effect modifiers (Objective 2). By employing the DR-learner within the WATCH workflow, we aim to provide a nonparametric solution for assessing heterogeneity, suitable for both continuous and binary endpoints. % This approach will enable us to gain deeper insights into the variability of treatment effects across different subgroups, ultimately enhancing the precision and efficacy of treatments.

The remainder of the article is structured as follows: In Section \ref{sec:methods:treatment_effect}, we provide a brief review of methods for estimating treatment effects. The core of our novelty lies in Section \ref{sec:methods:DR_learner_WATCH}, where we implement the DR-learner and demonstrate how it can be used to assess heterogeneity under the three objectives of the WATCH framework. Section \ref{sec:simulations} presents a comprehensive study evaluating the performance of the suggested methods against other approaches in various simulated scenarios. Section \ref{sec:clinical_trial} showcases a real clinical case scenario, where we apply the suggested methods to assess heterogeneity and identify effect modifiers in psoriatic arthritis trials. Finally, Section \ref{sec:conclusions} concludes the article with a discussion and future directions.

\section{Methods}
\label{sec:methods}
\subsection{Background on estimating treatment effect}
\label{sec:methods:treatment_effect}
We assume access to an i.i.d. sample of observations of \(Z = (X, A, Y)\), where \(X \in \mathbb{R}^p\) are covariates, \(A \in \{0, 1\}\) is a binary treatment or exposure, and \(Y \in \mathbb{R}\) is an outcome of interest. Our work is applicable to randomized clinical trials (RCTs), where patients are randomly assigned to treatments, but also to observational data under standard assumptions, where the treatment is assigned based on observed characteristics. We denote the \(n\) realizations of these observations as \(\{(\mathbf{x}_i, a_i, y_i)\}_{i=1}^n\). % We assume the outcome to be continuous, but, as we will discuss later, our methods can be directly used or extended to hold for other types of outcome variables. 
Furthermore, it is useful to define the following nuisance functions: $\mu_a (\mathbf{x}) = \mathbb{E} \left(Y | X = \mathbf{x}, A = a \right),$ and   $\pi(\mathbf{x}) =  \mathbb{P}(A = 1 | X = \mathbf{x}).$

In this work, we will follow the potential outcome (or counterfactual) framework, a common approach for formalising causal inference introduced by J. Neyman\cite{Neyman1923} in the context of RCT. It is also known as the ``Rubin causal model'', since Rubin\cite{Rubin1974} extended the potential outcome framework to perform causal inference in both observational and experimental studies (for the curious reader, Rubin\cite{rubin2019essential} provides a review on the history of causal inference). Under this framework, we denote as $Y(1)$ the potential outcome that it would have been observed if the patient was receiving the treatment $A=1,$ while as $Y(0)$ if the patient received the treatment $A=0$. The individual treatment effect is defined by the difference: $Y(1) - Y(0).$  The fundamental difficulty of causal inference is that we observe only one of the potential outcomes for each patient, denoted by \( Y \), since each patient received either treatment \( A = 1 \) or control \( A = 0 \), and as a result the individual treatment effects are not observed.

\subsubsection{Average Treatment Effect}
The Average Treatment Effect (ATE) is a critical concept for quantifying the impact of a treatment.\cite{imbens2004nonparametric} It measures the difference in expected outcomes between treated and untreated units and is formally defined as $\text{ATE} := \mathbb{E}[Y(1) - Y(0)]$. { Practically, in RCT setting, one of the simplest ways to estimate the ATE is by taking the difference in the average observed outcomes between treated and control groups:
$\widehat{\text{ATE}}_{\text{DiffMeans}} = \overline{Y}_1 - \overline{Y}_0,$
where $\overline{Y}_1$ and $\overline{Y}_0$ denote the sample means of observed outcomes in the treated ($A=1$) and control ($A=0$) groups, respectively. This approach is straightforward and intuitive, relying solely on basic descriptive statistics. Another simple but more flexible approach involves regression-based modeling. By adjusting for observed covariates, regression models can help reduce the variance of ATE estimates. In a more general form, the following ordinary least squares (OLS) 
regression can be used: $
Y_i = \beta_0 + \beta_1 A_i + \beta_2^\top X_i + \epsilon_i,$ where $Y_i$ is the observed outcome, $A_i$ is the treatment indicator (1 for treated and 0 for control), $X_i$ is a vector of observed covariates, and $\beta_1$ provides an estimate of the ATE, which is unbiased in an RCT. In the special case where no covariates are  included ($X_i$ omitted), the OLS estimate of $\beta_1$ is algebraically identical to the unadjusted difference in mean outcomes between the treated and control groups.

While the estimators described so far are more traditional tools for estimating average treatment effects, they can be interpreted as special cases of methods derived from the causal inference literature. This literature offers principled frameworks for estimating the ATE, with three of the most commonly used being G-computation, inverse probability weighting (IPW), and augmented IPW (AIPW): \cite{Glynn_Quinn_2010, ding2024first}
\begin{align}
\widehat{\text{ATE}}_{\text{G-comp}} & = \frac{1}{n} \sum_{i=1}^{n} \left\{\widehat{\mu}_1 (\mathbf{x}_i) - \widehat{\mu}_0 (\mathbf{x}_i) \right\},  \label{eq:G-comp} \\
\widehat{\text{ATE}}_{\text{IPW}}  & = \frac{1}{n} \sum_{i=1}^{n} 
\left\{ \left(  \frac{a_i}{\widehat{\pi}(\mathbf{x}_i)} - \frac{(1-a_i)}{1-\widehat{\pi}(\mathbf{x}_i)} \right)y_i \right\} = \frac{1}{n} \sum_{i=1}^{n} 
\left\{   \frac{a_i - \widehat{\pi}(\mathbf{x}_i)}{\widehat{\pi}(\mathbf{x}_i)\left(1-\widehat{\pi}(\mathbf{x}_i)\right)} y_i\right\},  \label{eq:IPW}\\
% \widehat{\text{ATE}}_{\text{DR:1}} & =  \frac{1}{n} \sum_{i=1}^{n} \left\{   \frac{a_i - \widehat{\pi}(\mathbf{x}_i)}{\widehat{\pi}(\mathbf{x}_i)\left(1-\widehat{\pi}(\mathbf{x}_i)\right)} \left(y_i - \widehat{\mu}_{a_i} (\mathbf{x}_i) \right) + \widehat{\mu}_1 (\mathbf{x}_i) - \widehat{\mu}_0 (\mathbf{x}_i) \right\} \label{eq:DR:1} \\
\widehat{\text{ATE}}_{\text{AIPW}} & =  \frac{1}{n} \sum_{i=1}^{n} 
\left\{   \frac{a_i - \widehat{\pi}(\mathbf{x}_i)}{\widehat{\pi}(\mathbf{x}_i)\left(1-\widehat{\pi}(\mathbf{x}_i)\right)} y_i  +
\left( \left(1 - \frac{a_i}{\widehat{\pi}(\mathbf{x}_i)} 
\right)\widehat{\mu}_1 (\mathbf{x}_i) - \left(1 - \frac{1-a_i}{1-\widehat{\pi}(\mathbf{x}_i)} 
\right)\widehat{\mu}_0(\mathbf{x}_i) \right)
\right\}, 
\label{eq:AIPW} \\
& =  \frac{1}{n} \sum_{i=1}^{n} \left\{   \frac{a_i - \widehat{\pi}(\mathbf{x}_i)}{\widehat{\pi}(\mathbf{x}_i)\left(1-\widehat{\pi}(\mathbf{x}_i)\right)} \left(y_i - \widehat{\mu}_{a_i} (\mathbf{x}_i) \right) + \widehat{\mu}_1 (\mathbf{x}_i) - \widehat{\mu}_0 (\mathbf{x}_i) \right\} 
\label{eq:AIPW_2} 
\end{align}
where $\widehat{\mu}_1(\mathbf{x}_i)$ is the estimated conditional expectation of the outcome given $\mathbf{x}_i$
within the treated group, i.e. $\widehat{\mathbb{E}} \left(Y | X = \mathbf{x}_i, A = 1 \right)$ and $\widehat{\mu}_0(\mathbf{x}_i)$ defined analogously for the control group. Furthermore, $\widehat{\pi}(\mathbf{x}_i)$ is the estimated propensity score, that is the estimated conditional probability
of treatment given $\mathbf{x}_i,$ i.e. $\widehat{\mathbb{E}} \left(A | X = \mathbf{x}_i\right).$

G-computation, Equation \eqref{eq:G-comp}, allows for a flexible modeling of the outcome based on covariates and treatment, which can lead to accurate estimates if the model for the outcome is specified correctly. Interestingly, the regression based estimate of the treatment effect discussed earlier, which is the coefficient of the treatment indicator 
in a linear outcome model, can be seen as a special case of the G-computation formula when the outcome model is correctly specified. In a linear additive model, predicting each individual’s potential outcomes under treatment and control and then averaging over the covariates reduces algebraically to the regression coefficient on the treatment indicator.  

Inverse Probability Weighting (IPW), Equation \eqref{eq:IPW}, adjusts for confounding by reweighting the observed outcomes, relying on the propensity score model. This creates a pseudo-population where the covariates are balanced across treatment groups, mimicking some properties of randomized experiments. Interestingly, in RCT, where the  propensity score is constant and known, the IPW weights become constant within treatment groups, and the weighted means equal the raw group means, making the IPW estimator numerically identical to the difference in means estimator discussed above. 

G-computation is sensitive to misspecification of the outcome model; if the model is incorrect, the resulting estimates can be biased. Conversely, IPW is sensitive to misspecification of the propensity score model and can be highly unstable in the presence of extreme propensity scores, leading to high variance in ATE estimates. Augmented Inverse Probability Weighting (AIPW) addresses these issues through a doubly robust approach that combines outcome modeling with IPW, yielding consistent estimates if either model is correctly specified and, as a result, offering greater protection against misspecification. % AIPW estimator remains consistent for the ATE if either the propensity score model or the outcome regression is misspecified but the other is properly specified. %It can be expressed in two equivalent forms, each providing a different perspective on how it adjusts for treatment effects.  
Equations \eqref{eq:AIPW} and \eqref{eq:AIPW_2} present two equivalent versions of the AIPW estimator (for completeness, proof of equivalence is provided in Appendix \ref{app:proof}).  The expression presented in Equation \eqref{eq:AIPW} can be interpreted as an adjustment of the IPW estimator, since it corrects it by incorporating the outcome models $\widehat{\mu}_{a_i}(\mathbf{x}_i)$, ensuring robustness against misspecification of either the outcome models or the propensity score model $\widehat{\pi}(\mathbf{x}_i)$. AIPW often results in more efficient and less biased estimates compared to either G-computation or IPW alone.\cite{Glynn_Quinn_2010} However, implementation of AIPW methods can be more complex, requiring careful specification and estimation of both the outcome model and the propensity score model. %, and if both models are misspecified, the estimates can still be biased.  % for example when using a Generalized Linear Model (GLM) with a canonical link function and intercept to estimate  $\mu_a (\mathbf{x}) = \mathbb{E} \left(Y | X = \mathbf{x}, A = a \right)$.

Furthermore, it can be shown that under the i.i.d. sampling model the asymptotically most efficient estimator for estimating ATE must be of the following form:\cite{kennedy2024semiparametric, hines2022demystifying} $\frac{1}{n}\sum_{i=1}^n {\mu}_1 (\mathbf{x}_i) - {\mu}_0 (\mathbf{x}_i) + \frac{a_i - {\pi}(\mathbf{x}_i)}{{\pi}(\mathbf{x}_i)\left(1-{\pi}(\mathbf{x}_i)\right)} \left(y_i - {\mu}_{a_i} (\mathbf{x}_i) \right)$. This result follows from the fact that the estimator is based on the efficient influence function for estimation of the ATE, and the asymptotic efficiency refers to the fact that the estimator is asymptotically unbiased and with smallest possible variance for estimating the semiparametric functional of the difference in outcome means (ATE).  In practice the nuisance functions $\mu_1(.)$, $\mu_0(.)$ and in non-randomized studies also $\pi(.)$ need to be estimated, and this estimator corresponds exactly to the AIPW estimator presented earlier in Equation \eqref{eq:AIPW_2}. As we mentioned above, a further interesting property of this estimator is that consistency is ensured if only one of the estimators (either $\mu_1(.),\mu_0(.)$ or $\pi(.)$) is correctly specified, this property of the estimator has sometimes been called double-robustness. In randomized clinical trials (which we primarily consider in this article) $\pi(.)$ is known, which implies that the estimator is guaranteed to be consistent. For introduction and more details to efficiency bounds, efficient influence functions and the double-robustness property of the estimator above please see \cite{kennedy2024semiparametric, hines2022demystifying}. 

%In summary, within the context of RCTs, where the propensity score is predetermined (such as 0.50 in a balanced 1:1 trial), the AIPW estimator remains consistent even if the outcome models are misspecified. Furthermore, when the outcome models are reasonably well estimated, the AIPW estimator can offer enhanced efficiency compared to the IPW estimator.
}

\subsubsection{Conditional Average Treatment Effect and Meta-learners}
\label{sec:methods:meta_learners}
Another important estimand is the CATE, which represents the expected difference in outcomes when subjected to a treatment versus no treatment, given specific covariate values $(\mathbf{x})$. Formally, CATE is defined as:
\begin{equation}
    \tau(\mathbf{x}) = \mathbb{E}[Y(1) - Y(0) \mid X = \mathbf{x}].
\end{equation}

Here, $\mathbf{x}$ represents the covariates of a patient, and $\mathbb{E}$ denotes the expectation operator, thus CATE is then an estimate of the treatment effects for all patients with the covariate values equal to $\mathbf{x}$.  While ATE provides a  general measure of the treatment effect averaged over the entire population, the CATE is important because it allows us to understand the effect of a treatment on a more personal level, taking into account specific characteristics or conditions of each individual. Following the literature,\cite{Curth2024,hoogland2021tutorial} we refer to the CATE as the individualized effect. However, this estimand is not truly individual - it pertains to subgroups of individuals who share the same covariate profile. A truly individual treatment effect is fundamentally unobservable, as only one potential outcome can be realized for any individual at a given time, as previously discussed.

Estimating CATE is a complex task, especially in the presence of high-dimensional data and potential confounders. To tackle this challenge, a range of approaches known as \textit{meta-learners} have been developed.\cite{kunzel2019metalearners, jacob2021cate, curth2021nonparametric, kennedy2023towards} Meta-learners provide flexible and robust frameworks for CATE estimation by leveraging machine learning models and combining them in different ways. The key idea behind meta-learners is to transform the problem into more manageable steps, often allowing the use of existing predictive models to estimate the treatment effects. In the literature, numerous meta-learners have been proposed; here, we focus on a representative set of them, that they have a direct correspondence/relationship with the ATE estimators we presented in the previous section. 

The first meta-learners that have been suggested in the literature, where the S and T-learner. \textbf{S-learner}\cite{kunzel2019metalearners} uses a single model to predict the outcome by including both the treatment indicator and the covariates as inputs. This method leverages a single predictive model that accounts for the treatment effect implicitly:
\[ \widehat{\tau}_{\text{S}} (\mathbf{x}_i) = \widehat{\mu} (\mathbf{x}_i,1) - \widehat{\mu} (\mathbf{x}_i,0),\]
where \(\widehat{\mu} (\mathbf{x},a)\) represents the predictive model that is trained with inputs from the combined space of covariates and the treatment indicator.  This learner, used to estimate CATE, can be seen as the corresponding method for G-computation to estimate ATE, Equation (\ref{eq:G-comp}). Another approach, similar to the S-learner, is the \textbf{T-learner}.\cite{kunzel2019metalearners} In contrast to using a single model that incorporates both the covariates and the treatment, the T-learner builds two separate models. One model is trained  the data from the treated group (where $a = 1$ ), and the other model from the control group (where $a = 0$ ):
\[ \widehat{\tau}_{\text{T}} (\mathbf{x}_i) = \widehat{\mu}_1(\mathbf{x}_i) - \widehat{\mu}_0(\mathbf{x}_i), \]
where \(\widehat{\mu}_1 (\mathbf{x})\) represents the model trained using only treated data, while \(\widehat{\mu}_0 (\mathbf{x})\) using control data. In the pharmaceutical statistics literature, the S- and T-learners were introduced under the name ``Virtual Twins'' in the seminal work of Foster et al.\cite{fost:2011}

After these relatively simple approaches, a plethora of meta-learners have been proposed in the literature. These meta-learners use machine learning models to estimate nuisance parameters (i.e., outcome models and propensity scores) and provide more accurate and robust estimates of the CATE. A special class of these meta-learners splits the process into two parts, and to present them we will use the notation of Jacob\cite{jacob2021cate}: in the first step, a pseudo-outcome, defined as $\psi$, is derived, while in the second step, the pseudo-outcomes are regressed on the covariates to obtain the final estimate of the CATE. In our work, we will focus on the following two meta-learners:
\begin{description}
    \item[F-learner or IPW-learner:] F-learner\cite{kunzel2019metalearners}  or Inverse Probability Weighting (IPW)-learner\cite{curth2021nonparametric} focuses on reweighting the observed outcomes to create a pseudo-population where the treatment assignment is effectively randomized. The idea is to generate a pseudo-outcome using the estimated nuisance function for the propensity score: 
\[
    \widehat{\psi}_{\text{IPW}} (\mathbf{x}_i) = \frac{a_i - \widehat{\pi}(\mathbf{x}_i)}{\widehat{\pi}(\mathbf{x}_i)\left(1-\widehat{\pi}(\mathbf{x}_i)\right)} y_i.
\]
To get CATE, the pseudo-outcome is regressed on the covariates, i.e. $\widehat{\tau}_{\text{IPW}} (\mathbf{x}_i) = E[\widehat{\psi}_{\text{IPW}} (\mathbf{x}_i) | X = \mathbf{x}_i].$   This learner, used to estimate CATE, can be seen as the corresponding method for IPW to estimate ATE, Equation (\ref{eq:IPW}).

    \item[DR-learner:] Doubly Robust (DR)-learner\cite{kennedy2023towards} combines both the outcome model and the inverse probability weighting approaches, and it achieves doubly robust property, meaning it can provide consistent estimates if either the outcome model or the propensity score model is correctly specified. Kennedy\cite{kennedy2023towards} and Curth and Van der Schaar\cite{curth2021nonparametric} have suggested two equivalent expressions of the pseudo-outcome DR-learner:
\begin{align} 
\widehat{\psi}_{\text{DR:1}} (\mathbf{x}_i, y_i) &= \frac{a_i - \widehat{\pi}(\mathbf{x}_i)}{\widehat{\pi}(\mathbf{x}_i)\left(1-\widehat{\pi}(\mathbf{x}_i)\right)} \left(y_i - \widehat{\mu}_{a_i} (\mathbf{x}_i) \right) + \widehat{\mu}_1 (\mathbf{x}_i) - \widehat{\mu}_0 (\mathbf{x}_i),~~\text{and} 
\label{eq:DR:1:CATE} \\
\widehat{\psi}_{\text{DR:2}} (\mathbf{x}_i, y_i) &= \frac{a_i - \widehat{\pi}(\mathbf{x}_i)}{\widehat{\pi}(\mathbf{x}_i)\left(1-\widehat{\pi}(\mathbf{x}_i)\right)} y_i  +
\left(1 - \frac{a_i}{\widehat{\pi}(\mathbf{x}_i)} 
\right)\widehat{\mu}_1 (\mathbf{x}_i) - \left(1 - \frac{1-a_i}{1-\widehat{\pi}(\mathbf{x}_i)} 
\right)\widehat{\mu}_0(\mathbf{x}_i). 
\label{eq:DR:2:CATE} 
\end{align}
The expression in Equation (\ref{eq:DR:1:CATE}) can be seen as providing correction to a simple outcome-model estimator, i.e. $ \widehat{\mu}_1 (\mathbf{x}_i) - \widehat{\mu}_0 (\mathbf{x}_i).$  The corresponding correction is an IPW applied to the residuals of the outcome-model estimate.\cite{abecassis2024prediction}
On the other side, Equation (\ref{eq:DR:2:CATE}) can be seen as providing correction to the IPW estimator, i.e. $ \frac{a_i - \widehat{\pi}(\mathbf{x}_i)}{\widehat{\pi}(\mathbf{x}_i)\left(1-\widehat{\pi}(\mathbf{x}_i)\right)} y_i$. The correction this time corresponds to an outcome
model reweighted by the residual treatment probabilities.\cite{abecassis2024prediction} In Appendix \ref{app:proof} we provide a detailed proof over the equivalence of these two expressions. The pseudo-outcome is regressed on the covariates to get the final estimate of $\widehat{\tau}_{\text{DR}} (\mathbf{x}_i) = E[\widehat{\psi}_\text{{DR}} (\mathbf{x}_i) | X = \mathbf{x}_i].$ This learner, used to estimate CATE, can be seen as the corresponding method for AIPW to estimate ATE, Equation (\ref{eq:AIPW}).
\end{description}

To sum up, these meta-learners employ diverse methodologies for estimating CATE, utilizing both outcome modeling and propensity score techniques to produce results that are both robust and flexible. One way to conceptualize these meta-learners for estimating CATE is to view them as a means of estimating the individual contributions (summands) that sum up to the overall estimation of ATE. These individual contributions can be seen as causal contrasts, $Y(1)-Y(0)$, in expectation.

\subsection{Using DR-learner to assess TEH within the WATCH framework}
\label{sec:methods:DR_learner_WATCH}
In the following sections, we will illustrate how the DR-learner procedure can be used to address the three objectives we aim to answer in WATCH to assess heterogeneity. However, before that, we will demonstrate how we implemented the DR-learner.

\subsubsection{Implementing DR-learner Using Stacking Machine Learning}
% \section{Constructing a DR-learner}
Kennedy \cite{kennedy2023towards} introduced the approach for constructing the DR-learner building upon the idea of estimating ``pseudo-outcomes'' as an intermediate step. Apart from providing a clear expression for the estimator (Equation \eqref{eq:DR:1:CATE}), Kennedy also introduced an algorithm (Algorithm \ref{alg:DR_learner}) that leverages cross-fitting techniques. Cross-fitting\cite{okasa2022meta} helps to mitigate overfitting and improve the estimator's performance by partitioning the data into folds. %Specifically, one subset of the data is used for estimating the propensity scores and outcome models, while another subset is used to perform the final regression to derive the CATE estimates. This careful division ensures that the estimation process does not overfit to any single part of the data, leading to more robust and generalizable results. Algorithm \ref{alg:DR_learner} provides a detailed description of the construction of the DR-learner.\citep{kennedy2023towards} This procedure involves splitting the dataset into multiple folds. 
The propensity score model \(\widehat{\pi}(x)\) and the outcome models \(\widehat{\mu}_1(x)\) and \(\widehat{\mu}_0(x)\) are estimated using training data, then the  pseudo-outcomes are computed based on the estimated models. Finally, these pseudo-outcomes are used to estimate the CATE in the test data. At the end, the results from all folds are aggregated to produce the final doubly robust estimate of CATE. 

Kennedy's approach provides desirable guarantees, such as robustness to model misspecification - ensuring consistency as long as either the propensity or outcome model is correctly specified - while also allowing faster rates for estimating the CATE even when the nuisance estimates converge at slower rates. These guarantees enhance the reliability of the DR-learner in practical applications, ensuring that it delivers robust and accurate estimates.

\begin{algorithm}
\caption{DR-learner}
\textbf{Input}: $Z = \{y_i, a_i, \mathbf{x}_i\}_{i=1}^{n}$
\begin{algorithmic}[1]
\State Split sample $Z$ into $K$ random subsets (folds)
\For{$k \in \{1, \ldots, K\}$}
    \State {assign} samples to train/test set: $S_{train} = Z \setminus S_k$ and $S_{test} = S_k$
    \State \textbf{Step 1: Nuisance training}   
    \State {learn}  nuisance functions: $\widehat{\pi}, \widehat{\mu}_0$ and $\widehat{\mu}_1$ in $S_{train}$
    \State \textbf{Step 2:  Pseudo-outcome regression}
    \State {estimate} pseudo-outcomes: $\widehat{\psi}_{\text{DR},k} (\mathbf{x}_i, y_i) = \frac{a_i - \widehat{\pi}(\mathbf{x}_i)}{\widehat{\pi}(\mathbf{x}_i)\left(1-\widehat{\pi}(\mathbf{x}_i)\right)} \left(y_i - \widehat{\mu}_{a_i} (\mathbf{x}_i) \right) + \widehat{\mu}_1 (\mathbf{x}_i) - \widehat{\mu}_0 (\mathbf{x}_i) $ in $i \in S_{test}$
    \State {train} final CATE model: $\widehat{\tau}_{\text{DR},k}$ in $S_{test}$
    \State {estimate} CATE:  
    $\widehat{\tau}_k(\mathbf{x}_i) = \widehat{\tau}_{\text{DR},k}(\mathbf{x}_i) $ for $\mathbf{x}_i \in Z$
\EndFor (\textbf{Step 3:  Cross-fitting} by repeating the process with different folds)
\State generate vector of pseudo-outcomes:  $\widehat{\psi}_{\text{DR}} = \{\hat{\psi}_{\text{DR,1}}, \hat{\psi}_{\text{DR,2}}, \ldots, \hat{\psi}_{\text{DR},k}\}$
\State generate vector of CATE estimates: $\widehat{\tau}_{\text{DR}}(\mathbf{x}_i) = \frac{\sum_{k=1}^{K}\widehat{\tau}_k(\mathbf{x}_i)}{K}$ for $\mathbf{x}_i \in Z$
\end{algorithmic}
\label{alg:DR_learner}
\end{algorithm}

The pseudo-outcomes within the DR-learner can be conceptualized as individual treatment effects because they represent the estimated effect of the treatment on the outcome for each individual, adjusting for covariates. This creates a pseudo-observation that reflects the causal effect of the treatment at the individual level, i.e. ITE, which is the difference between the potential outcomes $Y(1)-Y(0)$ for that individual.

Here we will provide more details on how we implemented the DR-learner. In our work, we use 5-folds for the cross-fitting ($K=5$). As we see in Algorithm \ref{alg:DR_learner}, to construct the DR-learner we need to estimate in each fold three nuisance functions ($\widehat{\pi}, \widehat{\mu}_0, \widehat{\mu}_1$) and CATE ($\widehat{\tau}_{\text{DR}}$). To estimate the nuisance parameters we will utilize  flexible regression models using machine learning and implement a nested cross-validation process.  In our work we implement an ensemble of models using the \pkg{SuperLearner} R package,  an algorithm that employs cross-validation to estimate the performance of multiple models and then constructs an optimal weighted ensemble (stacking) based on their performance on the test sets within the inner folds of a nested cross-validation process.\citep{van2007super} Separate Super Learner models are employed to estimate the outcome functions for the control and treatment groups, $\widehat{\mu}_0, \widehat{\mu}_1$ respectively, the propensity  $\widehat{\pi}$ and CATE $\widehat{\tau}_{\text{DR}}.$ The cross-validation of the \pkg{SuperLearner} is nested within the cross-fitting, and we use 10 folds (the default parameter) to estimate the weights for combining the base learners - this is the internal mechanism that evaluates each learner’s performance and determines the optimal convex combination.

{ In this ensemble we included two diverse models: LASSO regression model using the \pkg{glmnet} R package\cite{glmnet} and random forest of conditional inference trees (cforest) implemented in \pkg{party} R package \cite{hothorn2015package}. The cforest algorithm is an ensemble learning method that constructs multiple conditional inference trees as base learners. Unlike traditional random forests, cforest leverages unbiased recursive partitioning based on permutation tests, which helps mitigate variable selection bias. Each tree is trained on a different bootstrap sample of the data, and predictions are aggregated across the ensemble to enhance model stability and predictive accuracy. This approach yields robust estimates, particularly in settings with complex interactions or non-linear relationships among predictors. The hyper parameters of the models need to be predefined or selected through a tuning process. In our implementation, we used the default settings provided by the respective packages, and even this can lead to competitive results as we illustrate in the simulations. We acknowledge that model performance can potentially be improved through more sophisticated hyper parameter tuning strategies.}
\ \\
\noindent \textbf{Related works:} Apart from the meta-learners we have discussed so far to derive pseudo-outcomes, there is a plethora of methods that have been suggested in the literature. Some other examples include the X-learner\cite{kunzel2019metalearners}, which is an extension of the T-learner, specifically designed to handle scenarios with strong imbalances between treatment arms, while it is also adapt to handle situations with confounding variables. R-learner\cite{Nie2020} is a two-stage procedure where in the first stage the nuisance parameters are estimated, such as the propensity score and the outcome model. Residuals are then computed by subtracting the predicted values from the observed data and in the second stage, these residuals are used to construct a ``pseudo-outcome'',which is then modeled in order to estimate the causal parameter of interest. This two-step process enables the R-learner to accurately estimate the treatment effect by controlling for confounding variables. More recently, 
Efficient Plug-in (EP)-learner was introduced, \cite{van2024combining} which is based on a novel efficient plug-in estimator for the class of population risk functions under consideration, addressing challenges such as the instability associated with inverse probability weighting and the violation of bounds in pseudo-outcomes.

\subsubsection{Using DR-learner to assess the evidence against treatment effect homogeneity (Objective 1)}
\label{sec:DR_learner:objective_1}
In this section we will show how to use pseudo-outcomes to perform an overall test against homogeneity. We will utilize the conditional inference procedures leveraging a permutation test framework.\cite{hothorn2006lego} Our main aim is to test independence between the pseudo-outcomes \(\widehat{\psi}_{\text{DR}}\) and the set of covariates \(X = (X_1, X_2, \ldots, X_p)\). Formally, we test the null hypothesis $H_0: P(\widehat{\psi}_{\text{DR}} \mid X) = P(\widehat{\psi}_{\text{DR}})$ against arbitrary alternatives, where \(P\) denotes the probability measure.\footnote{Please note that in a formal mathematical perspective, rejecting this null does not prove that there is treatment effect heterogeneity. Indeed, treatment effects are heterogeneous precisely when the mean of $\widehat{\psi}_{\text{DR}}$ depends on $X$, but technically, it is possible that (e.g.) the variance of  $\widehat{\psi}_{\text{DR}}$  depends on $X$ but not the mean, in which case the null would be false while the treatment effect is homogeneous. However, our test statistics should only have power when $\mathbb{E}[\widehat{\psi}_{\text{DR}} \mid X]$ depends on $X$, so in practice this should not be an issue. Indeed, previous literature shows that permutation tests can often be robust to this sort of issue (see, e.g., Chung and Romano\citep{Chung2013}). Overall, our simulations show that the suggested methods effectively controls type I error while having high power.}

Permutation tests are one option for performing this test, and in our work we will consider the conditional inference framework reviewed in \cite{hothorn2006lego} and implemented in the \pkg{coin} R package\cite{hothorn2008implementing}. The default implementation uses linear statistics of the form 
\begin{equation} 
\mathbf{T}=\sum_{i=1}^N g(\mathbf{x}_i) \widehat{\psi}_{\text{DR}} (\mathbf{x}_i, y_i) \in \mathbb{R}^p,
\end{equation} 
where $g$ is a transformation of the covariates (this transformation is important when we have categorical variables), and $\widehat{\psi}_{\text{DR}} (\mathbf{x}_i, y_i)$ is the pseudo-outcome derived by the DR-learner for the { sample} $\mathbf{x}_i.$ 

{ To test for treatment effect heterogeneity, we construct a null distribution for the test statistic $\mathbf{T}$ by permuting the pseudo-outcomes relative to the covariates. This breaks any systematic association between the covariates and the pseudo-outcomes, as would be expected under the null hypothesis. Importantly, no additional holdout set is required for this testing procedure because we use cross-fitting when estimating the pseudo-outcomes. Cross-fitting ensures that the estimation of $\widehat{\psi}_{\text{DR}}(\mathbf{x}_i, y_i)$ for each observation is based on models trained on data that exclude that observation. This effectively mimics the role of a holdout set and prevents overfitting, thereby preserving the validity of the permutation test.}

The mean and covariance matrix under the null are then calculated to create a standardized version of $\mathbf{T}$. The test statistic $\mathbf{T}$ is a $p$-dimensional vector, i.e. $\mathbf{T} = \{T_1, ..., T_p\},$ and to reduce it to a univariate statistic (and thus perform a global test), Hothorn et al.\cite{hothorn2008implementing} suggest the maximum or the quadratic form of the standardized versions of $\mathbf{T}$ as the final test statistic. % Note that normal asymptotic approximations, based on \cite{strasser1999}, are also implemented in the coin R package, which we will use in the following sections.
 % Once the variables are appropriately transformed, the test statistic is computed. Under high-dimensional settings, two common forms of the test statistic are employed:\cite{hothorn2008implementing}
\begin{description}
    \item[Max-type statistic] takes the  maximum of the absolute values of the standardized test statistics, i.e. $T_{\max} = \max_{j=1, \ldots, p} \left| T_j \right|.$ This approach is useful for detecting the largest effect among multiple comparisons.
 \item[Quadratic-type statistic] takes a quadratic form of the test statistics, $ T_{\text{quad}} = \sum_{j=1}^{p} T_j^2.$ This approach offers a collective deviation of the observed data from the null hypothesis by summing the squared standardized deviations of all comparisons.
\end{description}
The choice between the maximum and quadratic statistics depends on the specific hypothesis and the nature of the expected associations in the data. The maximum statistic is ideal for scenarios where a single predictor is expected to have a large impact, while the quadratic statistic is suited for detecting distributed moderate effects across multiple predictors. %Both statistics are implemented in the \pkg{coin} package, providing flexible and comprehensive tools for conditional inference testing. This flexibility allows researchers to select the appropriate statistic based on their research questions and the underlying structure of their data. 

%To test the null hypothesis, a permutation procedure is applied. First, the observed test statistic \(T_{\text{obs}}\) is computed based on the original dataset. Then, permutations of the outcome variable \(Y\) are generated while maintaining the input variables \(X\) fixed. For each permutation, the test statistic \(T\) is recalculated. This process is repeated a large number of times (e.g., 1000 permutations) to create the distribution of the test statistic under the null hypothesis \(H_0\). The p-value is then calculated as the proportion of permutations where the test statistic is at least as extreme as the observed statistic:
%\[p = \frac{\sum_{b=1}^{B} \mathbb{I}(T_b \geq T_{\text{obs}})}{B},\]
%where \(T_b\) is the test statistic for the \(b\)-th permutation, \(\mathbb{I}\) is the indicator function, and \(B\) is the total number of permutations.

Given that \(X\) may be a mixture of continuous and categorical variables, the transformation process defined by $g$ becomes crucial. Continuous variables generally undergo rank transformation to make the test robust against outliers and non-normality. Each continuous variable \(X_j\) is transformed into its rank among all observations. For categorical variables, contrast coding is typically applied, converting categorical variables into a set of binary variables representing the different levels.

In conclusion, the independence test using the permutation-based approach with max-type or quadratic-type statistics provides a robust framework for assessing the relationship between a scalar outcome and a set of input covariates, even when the data include a mixture of continuous and categorical variables. %The flexibility of the method allows it to adapt to various data types and dependency structures, making it a powerful tool in modern statistical analysis. 
Furthermore, the \pkg{coin} package \citep{hothorn2008implementing} provides the option to use an asymptotic approximation of the conditional null distribution of the test statistics, allowing for the efficient computation of p-values. This enhances the practicality of the method, particularly for large datasets. In Section \ref{sec:sim:objective_1}, we compare these two approaches (max-type and quadratic-type), and the best method is then compared against other methods suggested in the literature in Section \ref{sec:sim:competing_methods}.

\ \\
\noindent \textbf{Related works:} 
In their tutorial, Lipkovich et al. \cite[Section 6.1]{Lipkovich2024} provide a comprehensive review of various strategies for assessing heterogeneity. Chernozhukov et al. \cite{chernozhukov2017generic} introduced the concept of the best linear projection (BLP) of a ML proxy for the CATE, denoted as $\widehat{\Delta}(X_i)$. The BLP is formulated through the following equation:
\[ Y_i - \widehat{m}^{-i}(X_i) = \alpha \overline{\Delta} \left( A_i - \hat{\pi}^{-i}(X_i) \right) + \beta \left( A_i - \hat{\pi}^{-i}(X_i) \right) \left( \widehat{\Delta}^{-i}(X_i) - \overline{\Delta} \right) \]
In this equation, $\hat{m}^{-i}(X_i)$ and $\hat{\pi}^{-i}(X_i)$ represent the use of leave-one-out cross-validation (i.e. the predictions for the outcome and for the propensity score model are generated by excluding the  $i$-th observation from the dataset when fitting the models), and $\overline{\Delta}$ is the average treatment effect. %The ``$-i$'' superscript indicates that the prediction for a given input, $X_i$, was made by a model trained without the $i$-th observation.
This technique is commonly employed to reduce bias and improve estimation accuracy. A coefficient $\beta > 0$ signifies the presence of heterogeneity in the treatment effect. The BLP framework allows for the incorporation of ML estimators of CATE into a linear regression model, providing a method for testing hypotheses regarding treatment effect heterogeneity. The \pkg{grf} R package for Causal Forest\cite{wager2018estimation} offers an omnibus test to identify heterogeneity in treatment effects, based on the best linear fit of the target estimand. This test utilizes predictions made on held-out data and is grounded in the previously discussed equation, and we compare against this method in our simulations. Finally, the concept of Group Average Treatment Effect (GATE) testing, introduced by Chernozhukov et al.\cite{chernozhukov2017generic} and further developed by Imai and Li\cite{ImaiLi2024}, plays a pivotal role in assessing heterogeneity in treatment effects across different subpopulations. The null hypothesis for GATE testing asserts that $E(\Delta(X)|G_1) = \cdots = E(\Delta(X)|G_K)$, where the groups $G_K$ are defined using a general ML technique for estimating CATE. Imai and Li's\cite{ImaiLi2024} significant contribution is their development of a cross-validation or cross-fitting framework, which enhances the robustness of testing for treatment effect heterogeneity. This framework is versatile and can be applied regardless of the specific ML algorithm used for estimating CATE. Additionally, they derived the asymptotic variance for the test statistics within this cross-fitting framework, providing a solid foundation for evaluating the homogeneity of treatment effects. Finally, in a recent work Ji et al. \cite{Asher2024causalbounds} introduced dual bound method to estimate and perform inference on a class of partially identified causal parameters. { Interestingly, these bounds can be applied to the variance of the CATE to test for heterogeneity.} While the primary focus of this work is on estimation rather than testing, the derived bounds offer a valuable byproduct for testing purposes. %, and we also compare this method in the simulations (Section \ref{sec:sim:objective_1}).

\subsubsection{Using DR-learner to identify effect modifiers (Objective 2)}
\label{sec:DR_learner:objective_2}
% To answer the question on which covariates influence the treatment effect, or in other words which covariates act as effect modifiers, we will use again the pseudo-outcomes. To do that in a multi-variate fashion, we will use the data $(X, \widehat{\phi}) $ to build a regression model, and then we will rank the different covariates on their variable importance score. Because the clinical variables are of mixed data (continuous and categorical) it is important not to have any selection bias towards covariates with many possible splits. In our work we will use conditional inference trees,\cite{hothorn2006unbiased} an unbiased recursive partitioning method, and more specifically we will use the \pkg{party} R package\cite{hothorn2015package} to build conditional random forests, a forest based ensemble that utilizes conditional inference trees as base learners. Finally, to derive the variable importance scores we are using \pkg{permimp }\cite{debeer2020conditional} to calculate the permutation based importance for each covariate. These scores capture how ``important'' each covariate is in predicting the pseudo-outcomes, and the higher the score the more important the covaraite. The top covariates are the most promising effect modifiers to explore further. 

To determine which covariates influence the treatment effect, we will derive importance scores from the pseudo-outcomes and rank the variables based on their impact on the treatment effect. To accomplish this in a multivariate fashion, we will use the data $(X, \widehat{\psi}_{\text{DR}})$, where $X$ represents the set of covariates and $\widehat{\psi}_{\text{DR}}$ represents the pseudo-outcomes derived in Line 11 of Algorithm \ref{alg:DR_learner}, to build a regression model. Then, to rank the different covariates based on their variable importance scores, we employ a method that ensures unbiased selection \cite{strobl2007bias}, which is particularly important given that our clinical variables include a mix of continuous and categorical data types. Random forest for example is known to exhibit a bias toward variables that offer more potential split points — such as continuous variables or categorical variables with many levels — because these variables are more likely to produce splits that reduce impurity. 

In our work, we use conditional inference trees, an unbiased recursive partitioning method specifically designed to handle such biases.\cite{hothorn2006unbiased} %Conditional inference trees separate the variable selection process from the split point determination, which significantly reduces bias and enhances the validity of the importance scores. 
Conditional inference trees are a type of decision tree algorithm that addresses the selection bias inherent in traditional decision trees by using statistical tests to select splits, rather than choosing splits based solely on impurity reduction measures. This approach leads to more unbiased variable selection and often more reliable importance scores, particularly when dealing with variables of differing types. More specifically, we will utilize the \pkg{party} R package \cite{hothorn2015package} to construct a random forest of conditional inference trees (cforest). The main way to derive variable importance scores in the context of cforest is the permutation importance. This approach involves shuffling the values of a predictor variable and then measuring the subsequent decrease in model accuracy. The underlying intuition is straightforward: if a variable is crucial to the model's predictive power, randomly permuting its values will significantly disrupt the model's performance, leading to a notable decline in the performance. It reflects the contribution of each variable to the model's predictive power, and it is  useful when the goal is to identify which variables have the largest direct impact on model performance. To derive these importance scores, we will then use the \pkg{permimp} R package \cite{debeer2020conditional}. In our case this importance score is an interpretable metric that capture how ``important'' each covariate is in predicting the pseudo-outcomes. Covariates with higher importance scores are more influential in determining the treatment effect and are thus identified as potential effect modifiers. The top-ranking covariates are chosen for further exploration in subsequent analyses. 

The approach we followed in this section, which involves deriving pseudo-outcomes and then using a method to obtain variable importance scores by regressing $X$ on these pseudo-outcomes, is highly versatile. For example, this approach can be applied with any other method for deriving importance scores from pseudo-outcomes, such as those discussed by Hooker et al.\cite{hooker2021unrestricted}  In the simulations (Section \ref{sec:sim:objective_2}), we compare against SHAP (SHapley Additive exPlanations) values, which is another state-of-the-art method to derive variable importance scores.\citep{Lundberg2017}  %In the simulations (Section \ref{sec:sim:objective_2}), we compare against SHAP (SHapley Additive exPlanations) values, which is another state-of-the-art method based on cooperative game theory to provide a consistent and fair measure of feature importance. SHAP values consider the contribution of each feature to every possible subset of features, ensuring that the importance scores are equitably distributed among all predictors.\citep{Lundberg2017} 
% In Section \ref{sec:sim:objective_2}, we compare these two approaches. The best method is then compared against other methods suggested in the literature in Section \ref{sec:sim:competing_methods}.
\ \\
\noindent \textbf{Related works:}  Importance scores that capture how strongly a variable modifies treatment effects can also be derived from Causal forest,\cite{wager2018estimation} and in our simulation study, in the simulations we compare against this method in Section \ref{sec:sim:competing_methods}. Other works have developed methods to formally test which covariates modify treatment effects,\cite{Sechidis2021, hines2022variable, paillard2024} but this alternative objective is beyond the scope of this work---we instead focus on Objective 2, getting a good (unbiased) ranking of which covariates modify the treatment.

\subsubsection{Using DR-learner to estimate individualized treatment effects (Objective 3)}
\label{sec:DR_learner:objective_3}
DR-learner by design is tailored in estimating CATE (this outcome is provided in Line 12 of Algorithm \ref{alg:DR_learner}).\cite{kennedy2023towards}  One of the key features of the DR-learner is its use of cross-fitting, a technique that plays a critical role in reducing bias and improving the precision of causal estimates. %Cross-fitting involves splitting the data into multiple folds and ensuring that the estimation of nuisance parameters, such as the propensity score model and outcome models, does not use information from the same fold in which the causal effect is estimated, i.e. training the model that estimates CATE.\cite{okasa2022meta}  By implementing cross-fitting, the DR-learner reduces the potential for bias that could arise from using the same sample both to estimate nuisance parameters and to compute the treatment effects, and as a result it helps to mitigate the risk of overfitting, yielding more reliable estimates of the treatment effects.
While Kennedy\cite{kennedy2023towards} suggested one way of performing cross-fitting in DR-learner, various other alternatives can be performed. For example, Jacob\cite[Algorithm 2]{jacob2021cate} uses an alternative implementation, which firstly derives the pseudo-outcomes for all the examples using cross-validation, and in a second layer of cross-fitting CATE is estimated through an out-of-bag procedure. Another implementation of cross-fitting is suggested by Jacob\cite[Algorithm 1]{jacob2020cross}, which can be seen as the bootstrapped version of the original algorithm suggested by Kennedy\cite{kennedy2023towards}. In our work, we will explore two different variations of the cross-fitting:
\begin{description}
    \item[CATE from DR-learner:] In this approach we are using CATE estimates returned by the original methodology, in other words we are using the $\widehat{\tau}_{\text{DR}}$ estimates from Line 12 of Algorithm \ref{alg:DR_learner}.
    \item[CATE from OOB estimates of cforest:] In this approach we utilise the conditional random forest  model that we build using  $(X, \widehat{\psi}_{\text{DR}})$ to derive the effect modifiers for Objective 2 (Section \ref{sec:DR_learner:objective_2}), where $\widehat{\psi}_{\text{DR}}$ represents the pseudo-outcomes derived in Line 11 of Algorithm \ref{alg:DR_learner}. While we build this model to derive variable importance scores, we can use it also to predict CATE using the out-of-bag prediction capabilities that it offers. This approach shares similarities with the cross-fitting for DR-learner suggested by Jacob\cite[Algorithm 2]{jacob2021cate}.
\end{description}

In Section \ref{sec:sim:objective_3}, we compare these two approaches and the best method is then compared against other methods suggested in the literature in Section \ref{sec:sim:competing_methods}.
 
\section{Simulations study}
\label{sec:simulations}
In this section, we will perform a simulation study to compare all the different flavors of the DR-learner we described in Section 2. To generate scenarios that mimic clinical trial data, we will use the \pkg{benchtm} R package. Section \ref{sec:datagen} provides a brief description of how the scenarios were simulated, while Sun et al. \cite{sun2022comparing} provide more detailed information on these scenarios. Section \ref{sec:performance_measures} discusses the performance measures for assessing TEH under the three objectives of the WATCH framework. Sections \ref{sec:sim:objective_1}, \ref{sec:sim:objective_2}, and \ref{sec:sim:objective_3} present our results on using different flavors of the DR-learner for Objectives 1, 2, and 3, respectively. Finally, Section \ref{sec:sim:competing_methods} compares the suggested DR-learner against various competing methods.

\subsection{Data generation scenarios}
\label{sec:datagen}
Let $Y$ represent a clinical endpoint, $\textbf{X} = (X_1, X_2, \ldots, X_p)$ denote the vector of $p$ baseline covariates, and $A \in \{0,1\}$ the binary treatment indicator.  We set $P(A = 1|X = \textbf{X}) = P(A = 0|X = \textbf{X}) = 0.5$. The response is generated from
\begin{equation}
\label{form:data-generation}
f(\textbf{X}, A) = f_{\text{prog.}}(\textbf{X}) + A(\beta_0 + \beta_1 f_{\text{pred.}}(\textbf{X})),
\end{equation}
where $f_{\text{prog.}}(\textbf{X})$ represents the prognostic function and $f_{\text{pred.}}(\textbf{X})$ represents the predictive function. In our simulation study, we focus on continuous response variables, $Y = f(\textbf{X}, A) + \epsilon$, where $\epsilon \sim N(0, 1)$, although our methods also apply to binary responses and the real case study in Section \ref{sec:clinical_trial} focuses on this endpoint type. We use $p=30$ baseline covariates, which typically include demographic, disease severity or subtype covariates, and drug mechanism-related biomarkers. This number is realistic based on our experience with clinical studies. Four scenarios for $f_{\text{prog.}}(\textbf{X})$ and $f_{\text{pred.}}(\textbf{X})$ are shown in Table \ref{tab:sim_models}.

{ The parameters $\beta_0$ and $\beta_1$ are jointly chosen to achieve a power of 0.5 for an unadjusted Gauss test of the overall treatment effect. This reflects a realistic scenario in a Phase III study, where the observed treatment effect may be slightly smaller than what was anticipated at the design stage. In such cases, there is often increased interest in investigating biomarkers that may modify the treatment effect and identifying subgroups with potentially enhanced responses. The parameter $\beta_1$ plays a central role in determining the ``difficulty'' of a simulation scenario, as it governs the variability of treatment effects across the population. Setting this parameter realistically is challenging, for that reason given the limited empirical guidance, we explore a range of $\beta_1$ values in our simulations to capture different levels of heterogeneity (a zero value indicates identical treatment effects across patients, while a large positive or negative value suggests that treatment effects vary according to $f_{\text{pred.}}(\textbf{X})$). Values $\beta^*_1$ are calculated to achieve a power of 0.8 for testing $H_0: \beta_1=0$ versus $H_1: \beta_1 \neq 0$ given the data-generating model in \eqref{form:data-generation}, with a Type I error rate of 0.1. Scenarios $0,0.5\beta^*_1,\beta^*_1,1.5\beta^*_1,2\beta^*_1$ are considered for $\beta_1$. For scenarios where $\beta_1 \geq \beta^*_1$, treatment effect modification is realistically detectable, assuming known $f_{\text{prog.}}(\textbf{X})$ and $f_{\text{pred.}}(\textbf{X})$. The value $\beta_0$ is chosen to achieve an overall treatment effect power of 0.5 in each scenario. Refer to Sun et al.\cite{sun2022comparing} for details on the analytical power calculation.}

\begin{table}[tb]
\centering
\caption{Simulation models. Here $a\lor b$ is a logical statement representing ``a or b'', $a\land b$ a logical statement representing ``a and b'', $I(.)$ is the indicator function, $\Phi$(.) is the cumulative distribution function (cdf) of standard normal distribution, and $s$ is a scaling factor that is chosen depending on the simulation scenario to achieve a specific $R^2$ on the control group ($A=0$). Each of the four scenarios is run for 5 settings (corresponding to $\beta_1=0,0.5\beta^*_1,\beta^*_1,$ $1.5\beta^*_1,2\beta^*_1$). } 
\label{tab:sim_models}
 \begin{tabular}{l l} 
 \hline
No. & Models for covariates from mimic real data \\ \hline
Scenario 1 & $f(\textbf{X}, A) =  s\times (0.5I(X_1=\text{Y})+X_{11}) + A(\beta_0 + \beta_1\Phi(20(X_{11}-0.5)))$ \\
Scenario 2 & $f(\textbf{X}, A) = s\times (X_{14}-I(X_8=\text{N})) + A(\beta_0 + \beta_1X_{14})$ \\ 
Scenario 3 & $f(\textbf{X}, A) = s\times (I(X_1=\text{N})-0.5X_{17}) + A(\beta_0 + \beta_1I((X_{14}>0.25)\land (X_1=\text{N})))$ \\ 
Scenario 4 & $f(\textbf{X}, A) = s\times (X_{11}-X_{14}) + A(\beta_0 + \beta_1I((X_{14}>0.3)\lor(X_4=\text{Y})))$ \\ \hline
\end{tabular}
\end{table}

Our simulation study also generates covariates to emulate the joint distribution of predictors from Phase III pharmaceutical trials for an inflammatory disease. This approach approximates both marginal distributions and predictor dependencies. We use 30 baseline covariates (8 categorical, 22 numerical), including demographic and disease-related variables. We utilize the \pkg{synthpop} R package with default settings to generate synthetic data.\citep{nowok2016synthpop} For confidentiality, covariates are labeled $X_1, X_2, \ldots, X_{30}$, with categorical variable levels ciphered and numeric variables scaled to $[0,1]$.  Using this generated synthetic data set as input, the \texttt{synthpop} function generates data sets of $n=500$ for each simulation.

The treatment effect varies as a step-like (Scenario 1) or linear (Scenario 2) function of a single continuous covariate. Although step and linear functions might seem unrealistic, they lie at the extremes of a plausible range of smooth monotonic functions. Scenarios 3 and 4 involve step functions defined by two covariates, with ``and'' or ``or'' structures, respectively. Each scenario includes a covariate that is prognostic but not predictive, alongside one that is both prognostic and predictive. To account for the magnitude of prognostic effects on outcomes, we select the scaling factor $s$ in Table \ref{tab:sim_models} to achieve realistic $R^2$ values on how well covariates predict outcomes in the control arm in the original data. LASSO regression on real data models the outcome based on covariates for the control group, leading to $R^2$ value of $R^2=0.32$. This value then guided the selection of $s$ in Table \ref{tab:sim_models}, ensuring  it is replicated in the control arm across scenarios. More details on the simulation process can be found in Sun et al.\cite{sun2022comparing}

\subsection{Performance measures}
\label{sec:performance_measures}
We will compare the methods with respect to their performance on the following measures:
\begin{description}
    \item[Objective 1:] The treatment effect heterogeneity assessment is conducted, and each method provides a corresponding p-value. These p-values should have two desirable properties:
    \begin{description}
        \item[(i)] when data are simulated with no treatment effect heterogeneity, i.e. $\beta_1=0$, the p-values should follow a uniform distribution in $[0,1],$ 
        \item[(ii)] when data are simulated with treatment effect heterogeneity, i.e., $\beta_1>0$, smaller p-values indicate a more powerful method.
    \end{description}
    \item[Objective 2:] To identify effect modifiers, we consider the most predictive biomarker reported from each method. Again, there are two desirable properties:
        \begin{description}
        \item[(i)] when data are simulated with no treatment effect heterogeneity, i.e. $\beta_1=0$, an unbiased method should select each biomarker as the most predictive one with equal probability. In this case, we reported the probability that each biomarker is selected as the most important predictive biomarker. 
        \item[(ii)] when data are simulated with treatment effect heterogeneity, i.e., $\beta_1>0$, the true predictive biomarkers should have higher probability to be selected. In this case, we reported the probability that the top selected biomarker is truly predictive. 
    \end{description}
    \item[Objective 3:] { Each method returns an estimate of the individualized treatment effect (i.e., CATE). We evaluate accuracy using the Mean Squared Error (MSE) between the estimated and true CATE. Specifically, the MSE is computed for each simulation as the mean over all individuals, and then averaged across all simulation replications. Lower MSE values indicate more accurate estimations. To gain deeper insights into estimator behavior, we also assess bias and variance. Additionally, we compute Somers' D rank correlation coefficient\cite{harrel2001regression}  to assess how well  the estimated rankings of CATE align with the true (oracle) rankings.}
\end{description}

\subsection{Evaluating approaches within the DR-learner framework}
\label{sec:sim:DR-learner}
As presented in Section \ref{sec:methods:DR_learner_WATCH}, the suggested DR-learner framework for assessing heterogeneity is quite versatile, allowing for various approaches to address the three objectives. In this simulation study, we will empirically identify the best approach for each objective.  In Section \ref{sec:sim:competing_methods}, we will compare the top-performing DR-learner with various other competing methods from the literature.

\subsubsection{Evaluation with respect to overall test against homogeneity (Objective 1)}
\label{sec:sim:objective_1}
In this section we will compare the two tests of the DR-learner approach we presented in Section \ref{sec:DR_learner:objective_1} with two more tests we reviewed in the same section. The four methods we compare are the following. 
\begin{description}[noitemsep,topsep=0pt]
\item[DR-Maximum Statistic:] Utilize the \pkg{coin} R package to conduct an independence test between the multivariate covariate space and the pseudo-outcome, employing the maximum statistic and derive the p-values.
\item[DR-Quadratic Statistic:] Utilize the \pkg{coin} R package to conduct an independence test between the multivariate covariate space and the pseudo-outcome, employing the quadratic statistic and derive the p-values.
% \item[Dual Bound Test:] Utilize the \pkg{dualbounds} Python package to estimate the bounds on the variance of CATE through dual bound methodology and determine p-values from these calculated bounds  and subsequently, derive the p-values.
% \item[GRF Calibration Test:] Utilize the \pkg{grf} R package for Causal Forest to execute an omnibus test for identifying heterogeneity in treatment effects, which is based on the optimal linear fit of the target estimand and derive the p-values.
\end{description}

\begin{figure}[!h]
  \begin{center}
    \includegraphics[width=0.99\textwidth]{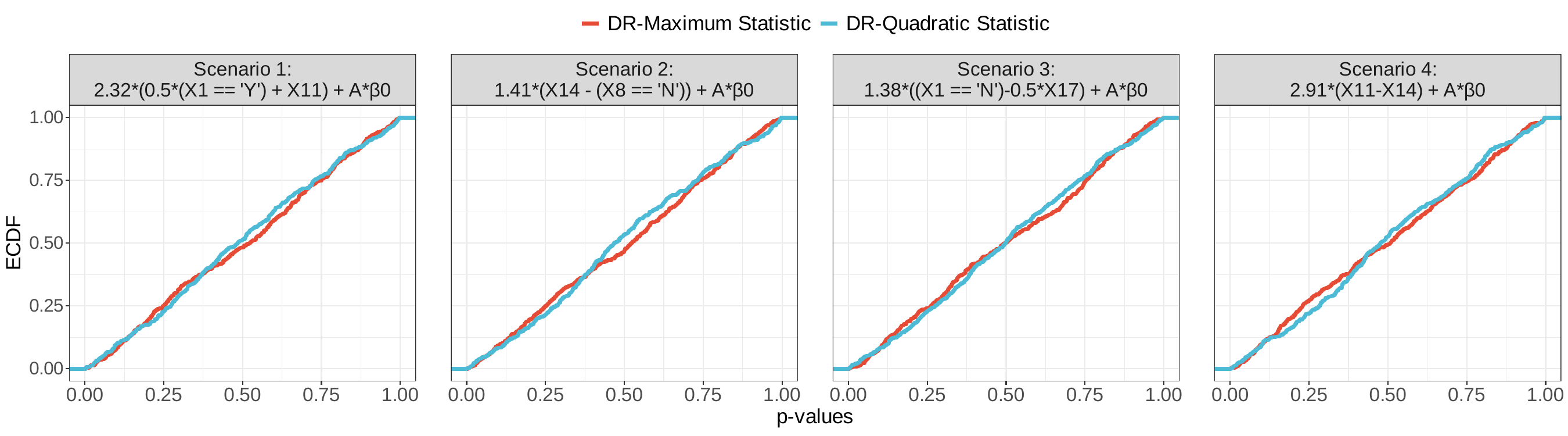}
  \end{center}
    \vspace{-0.5cm}
  \caption{\textbf{Comparison of the two methods for testing heterogeneity presented in Section \ref{sec:DR_learner:objective_1} with respect to Objective 1(i)}. Data are simulated under the condition of no treatment effect heterogeneity (i.e., $\beta_1 = 0$). { We present the ECDF of p-values under the null hypothesis. For uniformly distributed p-values, the ECDF follows a diagonal line. }}
  \label{fig:obj_1i}
\end{figure}

{ In Figure \ref{fig:obj_1i}, we present the empirical cumulative distribution function (ECDF) of p-values across 500 runs. Under the assumption of no treatment effect heterogeneity, the p-values are expected to follow a uniform distribution, resulting in an ECDF that aligns with the diagonal. Notably, this expected pattern is observed for both of our proposed methods.} %, whereas the other two methods show a skew towards high p-values.

\begin{figure}[!h]
  \begin{center}    \includegraphics[width=0.99\textwidth]{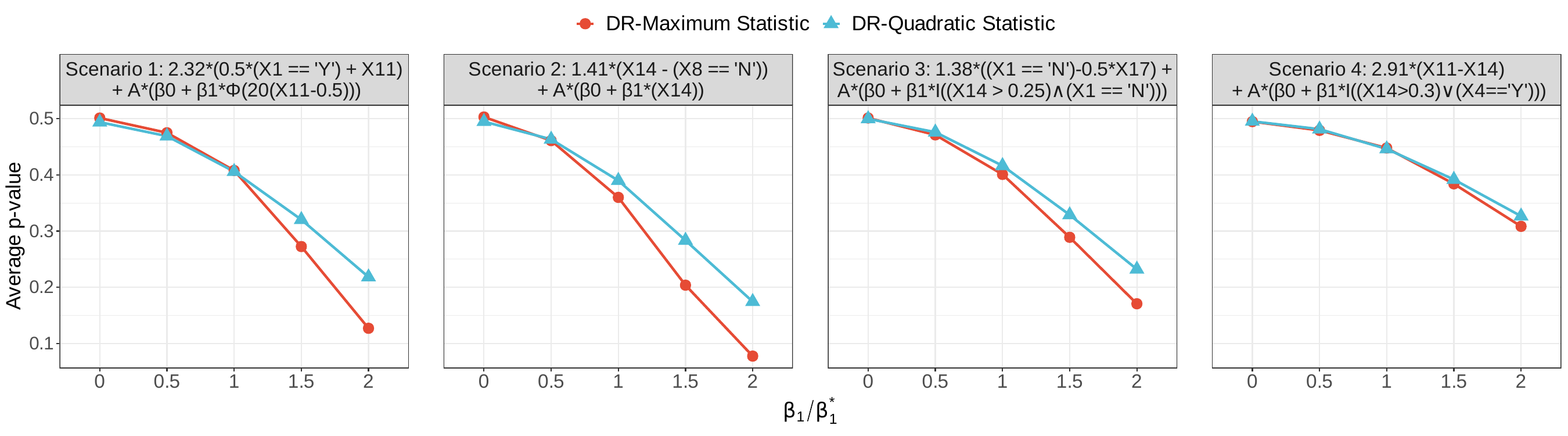}
  \end{center}
  %  \vspace{-0.75cm}
  \caption{\textbf{Comparison of the two methods for testing heterogeneity presented in Section \ref{sec:DR_learner:objective_1} with respect to Objective 1(ii)}. Data are simulated under various degrees of treatment effect heterogeneity (x-axis). We report the average p-values (across 500 runs) and when there is treatment effect heterogeneity (i.e., $\beta_1 > 0$), the lower the p-value, the more powerful the method.}
  \label{fig:obj_1ii}
\end{figure}

In Figure \ref{fig:obj_1ii}, we present the averages of the p-values over 500 runs in scenarios with various degrees of treatment effect heterogeneity (TEH). The lower the p-value, the more powerful the method.  Overall, the maximum statistic outperforms the quadratic statistic, and this trend is more pronounced when there is only one biomarker contributing to the TEH (for example Scenarios 1 and 2). The quadratic statistic may be more suitable for scenarios where many covariates interact to generate the TEH (for example Scenarios 3 and 4), as it focuses on the overall strength. In contrast, the maximum statistic may be more suitable when the effect modifiers do not interact to create the TEH.

\subsubsection{Evaluation with respect to identification of effect modifiers (Objective 2)}
\label{sec:sim:objective_2}
In this section, we compare two methods for deriving effect modifiers using the pseudo-outcomes from the DR-learner, as presented in Section \ref{sec:DR_learner:objective_2}. The methods we compare are the following. 
\begin{description}[noitemsep,topsep=0pt]
\item[DR-Perm(cforest):] Utilize the \pkg{party} R package to build a cforest (i.e. a forest of conditional inference trees) on the covariate space to predict pseudo-outcomes generated by the DR-learner. Then, use the \pkg{permimp} R package to derive permutation importance scores.
\item[DR-SHAP(rf):] Utilize the \pkg{ranger} R package to build a random forest on the covariate space to predict pseudo-outcomes generated by the DR-learner. Then, use the \pkg{treeshap} R package to derive TreeSHAP explanations and importance scores.
\end{description}

\begin{figure}[!h]
  \begin{center}
    \includegraphics[width=0.99\textwidth]{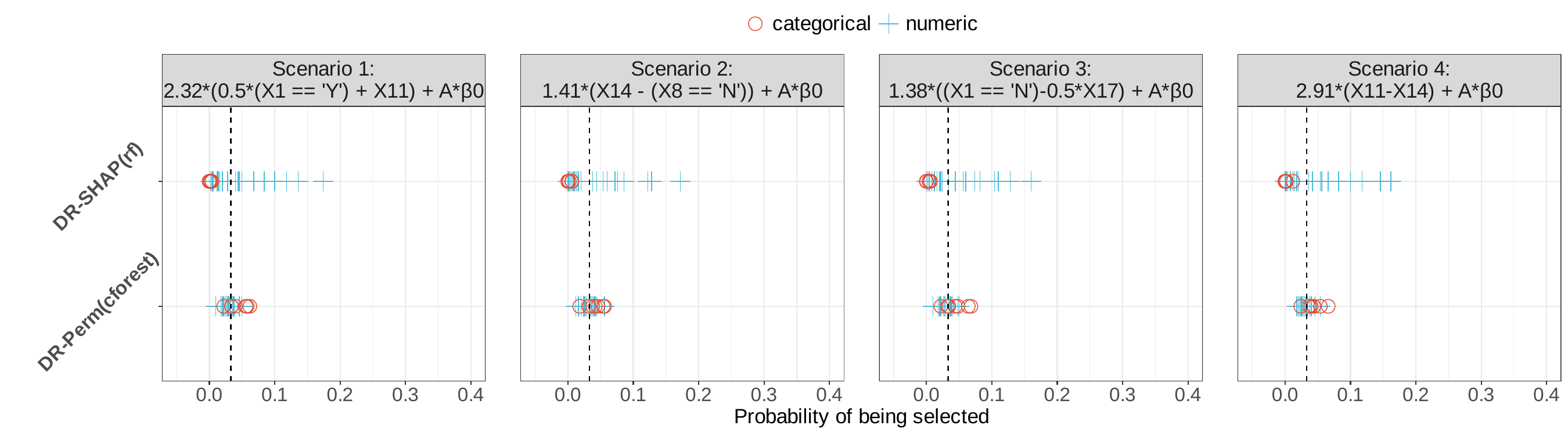}
  \end{center}
   % \vspace{-0.5cm}
  \caption{\textbf{Comparison of two methods for deriving effect modifiers (presented in Sec. \ref{sec:DR_learner:objective_2}) with respect to Objective 2(i)}. Data are simulated under the condition of no treatment effect heterogeneity (i.e., $\beta_1 = 0$). We report the average probability (across 500 runs) that each biomarker is selected as the most important predictive biomarker. Since there is no treatment effect heterogeneity, for a method to be unbiased all biomarkers should have probability equal to $1/30 \approx 0.03,$ dashed vertical line.}
  \label{fig:obj_2i}
\end{figure}

In Figure \ref{fig:obj_2i} we simulate scenarios without treatment effect heterogeneity, and we present how often each biomarker is selected as the most important. Under this scenario, a method without selection preference will have probability of selecting each biomarker close to $1/30 \approx 0.033$. The SHAP(rf) method is biased toward continuous biomarkers, since there are continuous biomarkers that have selection probability around 15\%, while all the categorical biomarkers have the smallest probabilities to be selected. 
{ The observed bias arises from the underlying random forest model used to compute SHAP values. As we discussed in Section~\ref{sec:DR_learner:objective_2}, RF models tend to favor continuous variables because they offer more potential split points, which can lead to greater impurity reduction during tree construction. This structural bias results in continuous variables being selected more frequently, thereby inflating their importance scores. When SHAP values are derived from such models, this bias is reflected in the attribution scores. Therefore, the apparent bias in SHAP(rf) is not a limitation of SHAP itself, but rather a consequence of the RF model’s internal preferences - a phenomenon that has been documented in the literature, e.g., Baudeu et al.\cite{biasSHAP2023}
}

\begin{figure}[!h]
  \begin{center}
    \includegraphics[width=0.99\textwidth]{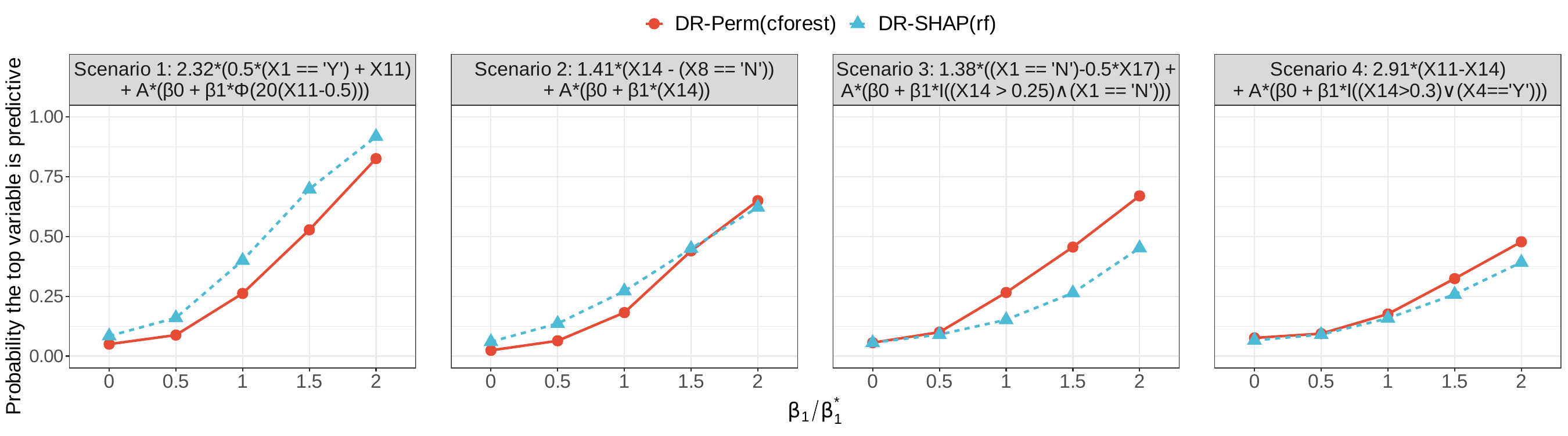}
  \end{center}
    \vspace{-0.5cm}
  \caption{\textbf{Comparison of two methods for deriving effect modifiers (presented in Sec. \ref{sec:DR_learner:objective_2}) with respect to Objective 2(ii)}. Data are simulated under various degrees of treatment effect heterogeneity (x-axis). We report the average (across 500 runs) probability that the top selected biomarker is truly predictive, and the higher this probability are the better the performance.}
  \label{fig:obj_2ii}
\end{figure}

In Figure \ref{fig:obj_2ii}, we present the probability that the top biomarker returned by each method to be an actual effect modifier across 500 runs in scenarios with various degrees of TEH. Both methods perform similarly, but we should mention that when biomarker interact to create TEH (Scenarios 3 and 4) the permutation approach outperforms SHAP in all settings. 

\subsubsection{Evaluation with respect to estimation of individualized treatment effects (Objective 3)}
\label{sec:sim:objective_3}
In this section we will compare the performance of various methods we discussed for estimating CATE. 

\begin{figure}[!h]
  \begin{center}
    \includegraphics[width=0.99\textwidth]{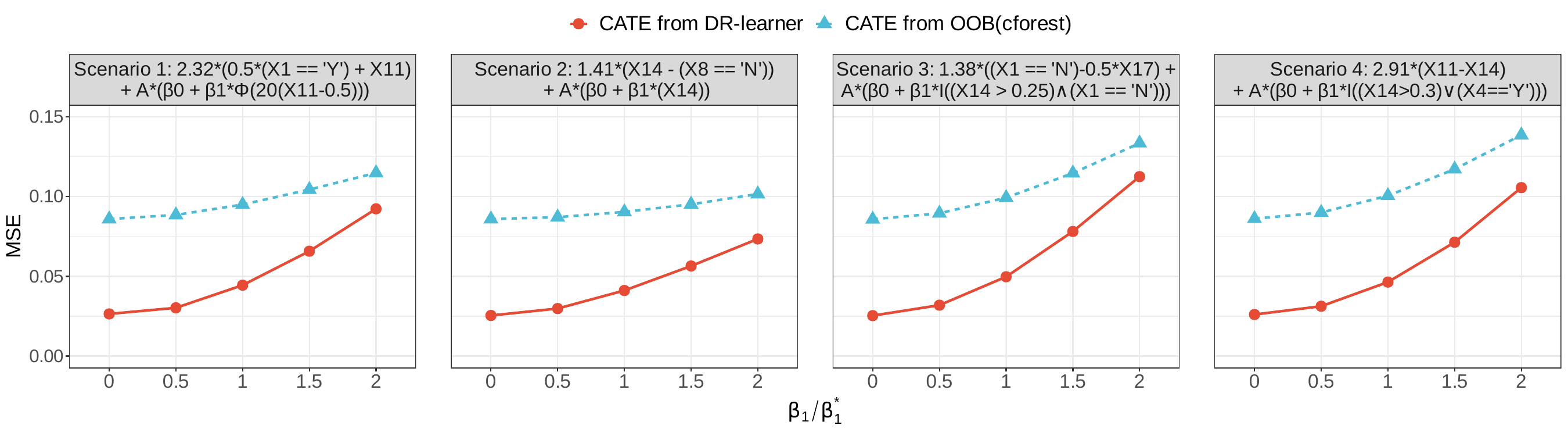}
  \end{center}
    \vspace{-0.5cm}
    \caption{\textbf{Comparison of two different cross-fitting strategies (presented in Sec. \ref{sec:DR_learner:objective_3}) with respect to Objective 3}. Data are simulated under various degrees of treatment effect heterogeneity (x-axis). We report the MSE (across 500 runs) for estimating CATE, and the lower this error is the better the performance of the cross-fitting strategy.}
  \label{fig:obj_3_cross_fitting}
\end{figure}

\paragraph{Comparison of different cross-fitting strategies}
Firstly, we will compare the two different cross-fitting approaches presented in Section \ref{sec:DR_learner:objective_3}.
\begin{description}[noitemsep,topsep=0pt]
\item[CATE from DR-learner:] Using the estimates of CATE returned by the original approach\cite{kennedy2023towards}, i.e. Algorithm \ref{alg:DR_learner}.
\item[CATE from OOB(cforest):] Using the out-of-bag predictions from the cforest we used for answering Objective 2, i.e. identify effect modifiers by building a cforest to predict the pseudo-outcomes. As we already mentioned, this approach can be seen similar to the one presented by Jacob\cite[Algorithm 2]{jacob2021cate}.
\end{description}
Figure \ref{fig:obj_3_cross_fitting} compares the two methods in terms of the MSE, and as we see the original cross-fitting  performs the best. This phenomenon can be attributed to the way CATE from OOB(cforest) is structured. In this method, pseudo-outcomes are initially estimated using all available data, and then the cforest is constructed based on this same dataset. Effectively, this equates to each data point being used twice; a practice which can introduce a risk of over-fitting.

\begin{figure}[!h]
  \begin{center}
    \includegraphics[width=0.99\textwidth]{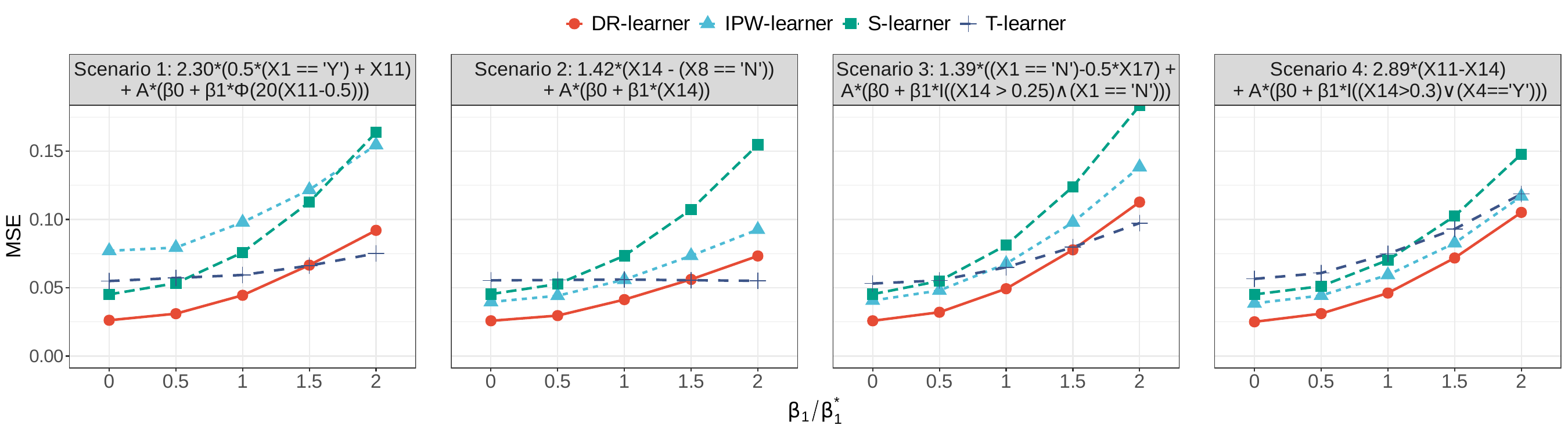}
  \end{center}
    \vspace{-0.5cm}
    \caption{\textbf{Comparison of four learners for estimating individualized treatment effect (presented in Sec. \ref{sec:methods:meta_learners}) with respect to Objective 3}. Data are simulated under various degrees of treatment effect heterogeneity (x-axis). We report the MSE (across 500 runs) for estimating CATE, and the lower this error is the better the performance of the learner.}
  \label{fig:obj_3_meta_learners}
\end{figure}

\paragraph{Comparison of different learners for estimating CATE}
{ Another insightful comparison is to evaluate the performance of the DR-learner against its ``main components'', the IPW and T/S-learner. Details about these learners can be found in Section \ref{sec:methods:meta_learners}. To keep the comparison fair, we will use the same nuisance models. Furthermore, for IPW and DR-learner we followed the same cross-fitting procedure described in Algorithm \ref{alg:DR_learner}. For the S/T-learner, no cross fitting is required, and we used data to learn the models. As shown in Figure \ref{fig:obj_3_meta_learners}, across all four scenarios, the DR-learner consistently achieves strong performance, yielding low MSE values across a wide range of signal strengths, and outperforming other methods in nearly all cases. This highlights its robustness and effectiveness in diverse settings.}

{
\paragraph{Advantages of doubly robust methods in the presence of model misspecification}
To further motivate the improvements offered by DR learner methods in RCTs, we conducted a focused simulation study that examines their performance under outcome model misspecification. Although RCTs eliminate confounding by design through randomized treatment assignment - making the propensity model trivially correct - the accuracy of singly robust learners (such as S/T-learner) still critically depends on correct specification of the outcome model, which, if misspecified, can lead to misleading treatment effect estimates even in experimental settings.

In our simulation study, we evaluated the performance of two learners, both relying on misspecified outcome models ($\mu_0$ and $\mu_1$): a single robust learner (T-learner) and a doubly robust learner (DR-learner). To simulate a scenario of severe misspecification, we fitted linear models for both outcomes using five variables—$X_{26}$, $X_{27}$, $X_{28}$, $X_{29}$, and $X_{30}$. As shown in Table \ref{tab:sim_models}, none of these variables were involved in the data-generating process. This setup mimics a situation in which we have collected five variables that are irrelevant to the endpoint and used them in a simple linear regression model. Such a design introduces substantial misspecification, making it particularly interesting to observe how the two learners behave under these conditions.

The results in Figure \ref{fig:obj_3_misspecified_mse} consistently showed that the DR learner achieved substantially lower MSE in all scenarios. This improvement was driven by a significant reduction in the variance (Figure \ref{fig:obj_3_misspecified_variance}) of the estimator, highlighting the stabilizing effect of the DR approach. In terms of bias (Figure \ref{fig:obj_3_misspecified_bias}),both learners performed similarly, with the doubly robust learner showing a slightly lower bias compared to the T-learner. This similarity can be attributed to the severe misspecification of the outcome models, which limits the ability of either approach to substantially reduce bias. 

The variance reduction observed suggests that even when the outcome model is poorly specified, the DR learner can maintain reliable estimation performance by effectively incorporating information from the treatment assignment mechanism. To sum it up, in the context of RCTs - as reflected in our simulated scenarios - where the propensity score is known by design, the DR learner emerges as a particularly effective approach for estimating individualized treatment effects.
\begin{figure}[!h]
  \begin{center}
    \includegraphics[width=0.97\textwidth]{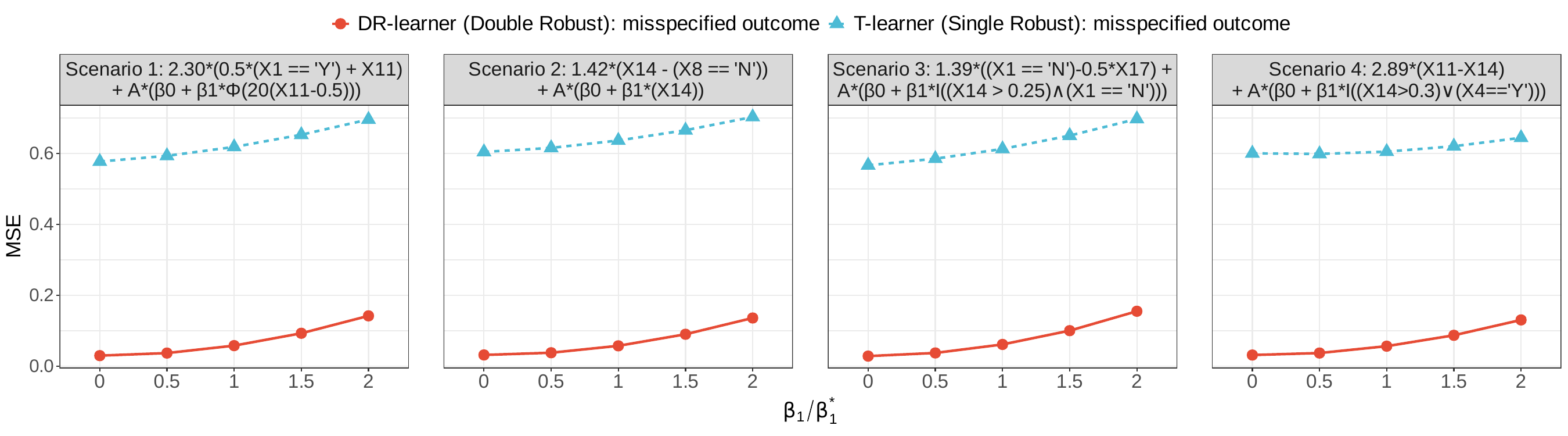}
  \end{center}
    \vspace{-0.5cm}
    \caption{\textbf{Comparison of single versus doubly robust learners for estimating individualized treatment effects under Objective 3, when the outcome model is severely misspecified, using MSE as evaluation metric}. Data are simulated under varying degrees of treatment effect heterogeneity (x-axis). For each run, we compute the mean squared error (MSE) of the estimated CATE: $ \frac{1}{n} \sum_{i=1}^{n} \left( \hat{\tau}_i - \tau_i \right)^2.$
    We then average this quantity across 500 runs. Lower values indicate more accurate and reliable estimators.}
  \label{fig:obj_3_misspecified_mse}
  \vspace{-0.25cm}
\end{figure}
\begin{figure}[!h]
  \begin{center}
    \includegraphics[width=0.97\textwidth]{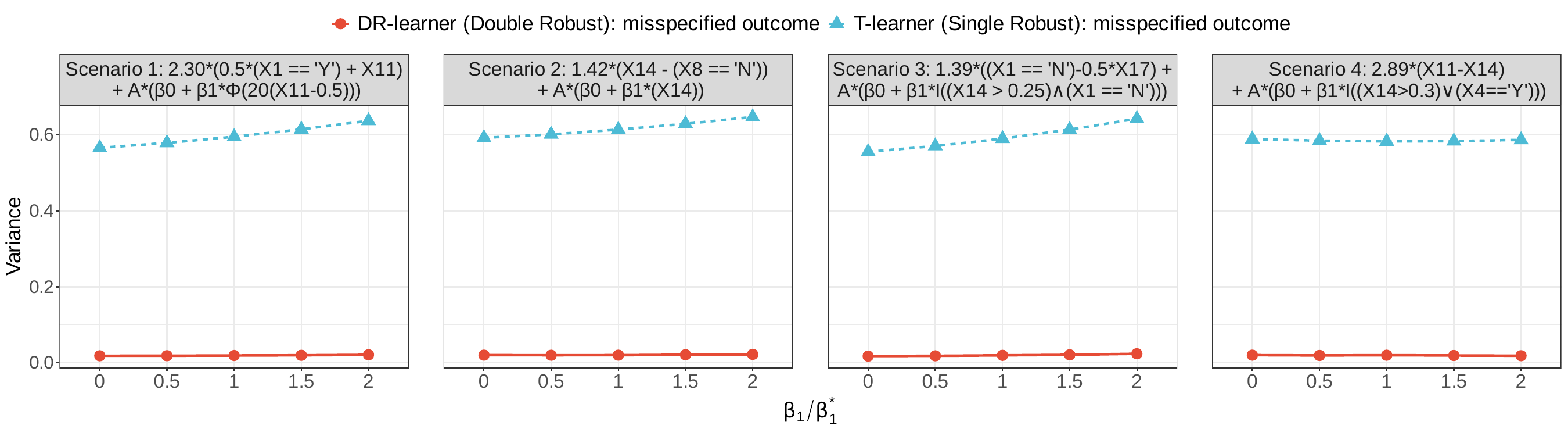}
  \end{center}
    \vspace{-0.5cm}
    \caption{\textbf{Comparison of single versus doubly robust learners for estimating individualized treatment effects under Objective 3, when the outcome model is severely misspecified, using variance as evalution metric}. Data are simulated under varying degrees of treatment effect heterogeneity (x-axis). For each run, we compute the variance of the estimated CATE across all examples:$
\frac{1}{n} \sum_{i=1}^{n} \left( \hat{\tau}_i - \bar{\hat{\tau}} \right)^2, \quad \text{where } \bar{\hat{\tau}} = \frac{1}{n} \sum_{i=1}^{n} \hat{\tau}_i.$ We then average this quantity across 500 runs. Lower values indicate more stable estimators.}
  \label{fig:obj_3_misspecified_variance}
    \vspace{-0.25cm}
\end{figure}
\begin{figure}[!h]
  \begin{center}
    \includegraphics[width=0.97\textwidth]{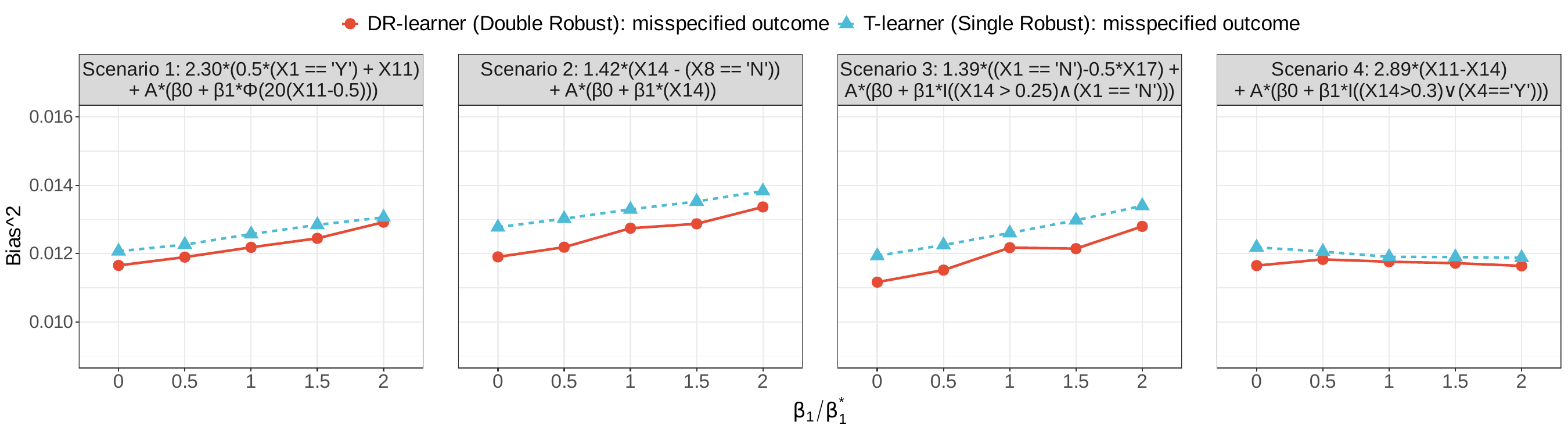}
  \end{center}
    \vspace{-0.5cm}
    \caption{\textbf{Comparison of single versus doubly robust learners for estimating individualized treatment effects under Objective 3, when the outcome model is severely misspecified, using bias as evaluation metric}. Data are simulated under varying degrees of treatment effect heterogeneity (x-axis). For each run, we compute the average bias of the estimated CATE:$
\frac{1}{n} \sum_{i=1}^{n} \left( \hat{\tau}_i - \tau_i \right).$ We then square this quantity and average it across 500 runs to obtain the empirical squared bias. Lower values indicate more accurate estimators.}
  \label{fig:obj_3_misspecified_bias}
\vspace{-0.25cm}
\end{figure}
}

% \subsubsection{Summary of optimal approaches within the DR-learner framework}
\subsection{Comparing our DR-learner proposal with various other competing methods}
\label{sec:sim:competing_methods}
In this section, we compare our DR-learner-based methodology for assessing heterogeneity with several competing approaches. We begin by describing each method under consideration. Next, we evaluate their performance across three key objectives, and finally, we summarize the results.
\ \\
\subsubsection{Describing competing methods}
{
We first present our proposed version of the DR-learner. We then describe several simple baseline methods, followed by more advanced approaches that build upon the concept of recursive partitioning
\paragraph{Our suggested method based on DR-learner}
Building on our results of Section \ref{sec:sim:DR-learner}, we will use the following implementation of the \textbf{DR-learner} to tackle the three objectives: 
\begin{description}
    \item[Objective 1:] Utilize the \pkg{coin} R package to conduct an independence test between the multivariate covariate space and the pseudo-outcome, employing the maximum statistic and derive the p-values.
    \item[Objective 2:] Utilize the \pkg{party} R package to build a cforest (i.e. a forest of conditional inference trees) on the covariate space to predict pseudo-outcomes generated by the DR-learner. Then, use the \pkg{permimp} R package to derive permutation importance scores.
    \item[Objective 3:] Using the estimates of CATE returned by the original approach\cite{kennedy2023towards}, i.e. Algorithm \ref{alg:DR_learner}.
\end{description}

\paragraph{Simple baseline methods}
To ensure a balanced evaluation, we include simple yet credible baseline methods for comparison alongside our more complex approach. These baseline has demonstrated good performance in a neutral benchmarking study conducted by Sun et al. \cite{sun2022comparing}.
\begin{description}
\item[Univariate:] A linear model is used, and a likelihood ratio test is performed for each biomarker \(X_j\) (\(j = 1, 2, \ldots, 30\)). This test compares the interaction model \(f(X_j, A) = \alpha_1 X_j + \beta_0 A + \beta_1 A X_j\) with the main effect model \(f(X_j, A) = \alpha_1 X_j + \beta_0 A\). The p-value \(p_j\) for each biomarker \(X_j\) is recorded.
\begin{description}
    \item[Objective 1:] We use \(\min(p_j) \times 30\) (Bonferroni adjustment) as the p-value for the treatment effect heterogeneity test.
    \item[Objective 2:] The biomarker \(X_j^*\) with the smallest p-value is selected as the top predictive biomarker.
    \item[Objective 3:] The model \(f(X_j, A) = \alpha_1 X_j^* + \beta_0 A + \beta_1 A X_j^*\), where \(X_j^*\) is the top predictive biomarker, is used to predict the treatment effect for each individual.
\end{description}

\item[Multivariate:] A linear model is used, and a likelihood ratio test is conducted to compare the interaction model 
\( f(\textbf{X}, A) = \sum_{j = 1}^{30}\alpha_j X_j + \beta_0 A + \sum_{j = 1}^{30}\beta_j A X_j \)
to the main effect model 
\( f(\textbf{X}, A) = \sum_{j = 1}^{30}\alpha_j X_j + \beta_0 A. \)
\begin{description}
    \item[Objective 1:] The corresponding p-value is recorded and used.
    \item[Objective 2:] The p-values from the interaction model for each interaction \(X_j A\) are compared, and the \(X_j^*\) with the smallest interaction p-value is selected as the top predictive biomarker.
    \item[Objective 3:] The interaction model 
    \( f(\textbf{X}, A) = \sum_{j = 1}^{30}\alpha_j X_j + \beta_0 A + \sum_{j = 1}^{30}\beta_j A X_j \)
    is used to predict the treatment effect for each individual.
\end{description}
\end{description}

\paragraph{Further methods build upon recursive partitioning}
Furthermore, our comparison also includes more advanced methods based on recursive partitioning, which have demonstrated competitive performance in the aforementioned neutral benchmarking study by Sun et al. \cite{sun2022comparing}.
\begin{description}
\item[MOB-L:] We use the R package \pkg{partykit} to implement the MOB method, utilizing the \texttt{glmtree} function for continuous outcomes. A variant of the MOB procedure is applied, where MOB is restricted to split only based on parameter instability for the treatment effect parameter in the model. This is achieved using the \texttt{parm} option (as described in the literature\citep{loh2019subgroup, thomas:2018}). Although this approach seems intuitive, Seibold et al.\citep{seib:zeil:hoth:2016} demonstrate that, in purely prognostic scenarios, it may lead to prognostic biomarkers being misidentified as predictive. This issue is also evident in the simulation results of Loh and Cao\citep{loh2019subgroup} and Thomas et al.\citep{thomas:2018} To address this, LASSO is first used to select prognostic biomarkers. This involves fitting LASSO regression models (\pkg{glmnet} package with the penalty parameter chosen by cross-validation and the \texttt{lambda.1se} option) 
\( f(\textbf{X}) = \sum_{j = 1}^{30}\alpha_j X_j \)
to each of the two treatment groups separately and selecting the union of the two sets of selected biomarkers \(X_1^*, X_2^*, \ldots, X_k^*\). The model fitted in each node in MOB is 
\( f(\textbf{X}) = \beta_0 A + \sum_{j = 1}^k \alpha_j X_j^* \)
where all biomarkers are used for node splitting selection. To build each tree, the model uses Bonferroni adjustment with \texttt{alpha = 0.10} and \texttt{minsize = 0.2 x 500 = 100}.
\begin{description}
    \item[Objective 1:] The Bonferroni adjusted p-value for each biomarker when testing the root node is extracted, and the smallest p-value is used.
    \item[Objective 2:] The biomarker with the smallest p-value is selected as the top predictive biomarker, regardless of its significance.
    \item[Objective 3:] Individualized treatment effects are predicted on a linear scale based on the model tree. Since a linear model with prognostic biomarkers is considered for each node, the MOB model is abbreviated as MOB-L.
\end{description}

\item[Causal Forest (CF):] To build the causal forest, we use 2000 trees (default parameter), and all tunable parameters are tuned by cross-validation using the \pkg{grf} package. 
\begin{description}
    \item[Objective 1:]  We directly report the p-value of the treatment effect heterogeneity test from the R package. The test is implemented with the \texttt{test\_calibration} function, which computes the best linear fit of the CATE effect using the forest prediction (on held-out data) and the mean forest prediction as the sole two regressors \citep{athey2019estimating}. This test for heterogeneity is motivated by the ``best linear predictor'' method of Chernozhukov et a.\citep{chernozhukov2017generic} 
    \item[Objective 2:] The most predictive biomarker is identified as the one with the largest variable importance score (a depth-weighted average of the number of splits on the biomarker of interest), calculated by the \texttt{variable\_importance} function.
    \item[Objective 3:] The individualized treatment effect (CATE) is returned by the model. 
\end{description}
\end{description}
}
% \item[Oracle] This approach is included as a benchmark since it reflects the best achievable performance if the true functional form were known. A linear or logistic regression model is fitted, including an intercept, the true prognostic \(f_{\text{prog.}}(\textbf{X})\), \(A\), and \(f_{\text{pred.}}(\textbf{X})\). The model is 
% \( f(\textbf{X}, A) = \alpha_0 f_{\text{prog.}}(\textbf{X}) + \beta_0 A + \beta_1 A f_{\text{pred.}}(\textbf{X}). \)
% \begin{description}
%    \item[Objective 1:] The treatment effect heterogeneity is assessed by testing whether the coefficient of $f_{\text{pred.}}(\textbf{X})$ is 0.
%    \item[Objective 3:] The predicted treatment effect is obtained from the fitted model.
% \end{description}

\subsubsection{Results Objective 1}
\begin{figure}[!h]
  \begin{center}
    \includegraphics[width=0.95\textwidth]{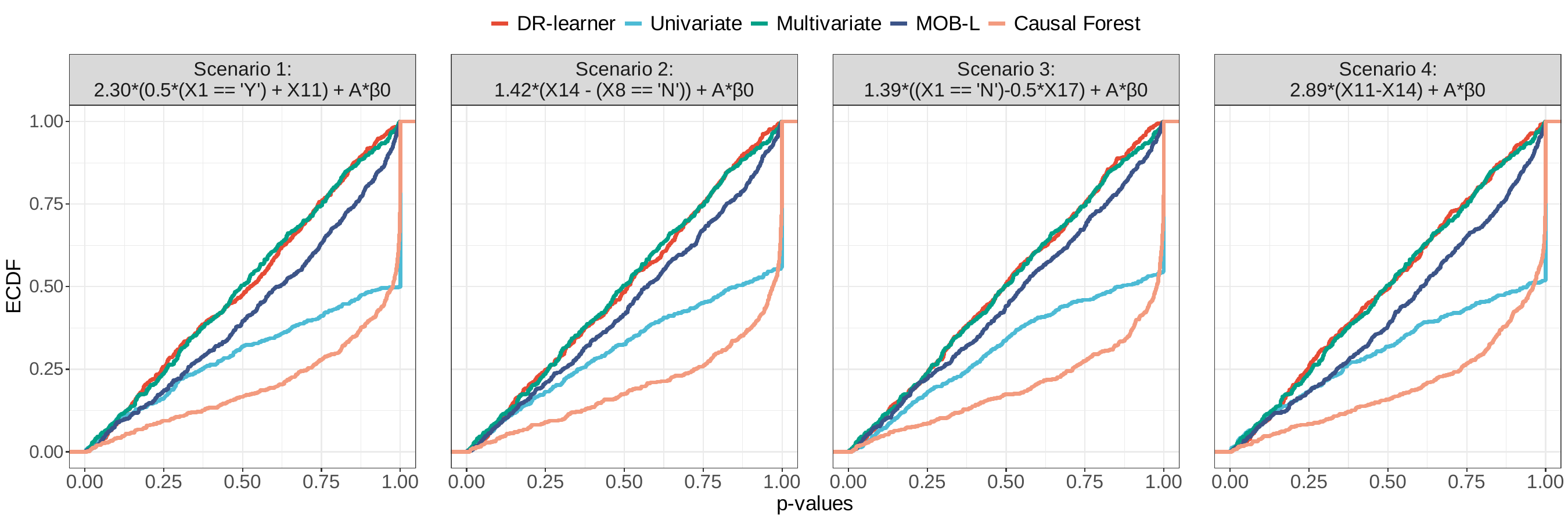}
  \end{center}
      \vspace{-0.5cm}
  \caption{\textbf{Comparison of our proposed DR-learner with competing methods regarding Objective 1(i)}. Data are simulated under the condition of no treatment effect heterogeneity (i.e., $\beta_1 = 0$). { We present the ECDF of p-values under the null hypothesis. For uniformly distributed p-values, the ECDF follows a diagonal line. When the ECDF curve lies below this diagonal, it indicates a higher proportion of large p-values.} }
  \label{fig:competing_methods_obj_1i}
\end{figure}

\begin{figure}[!h]
    \vspace{-0.5cm}
  \begin{center}
    \includegraphics[width=0.95\textwidth]{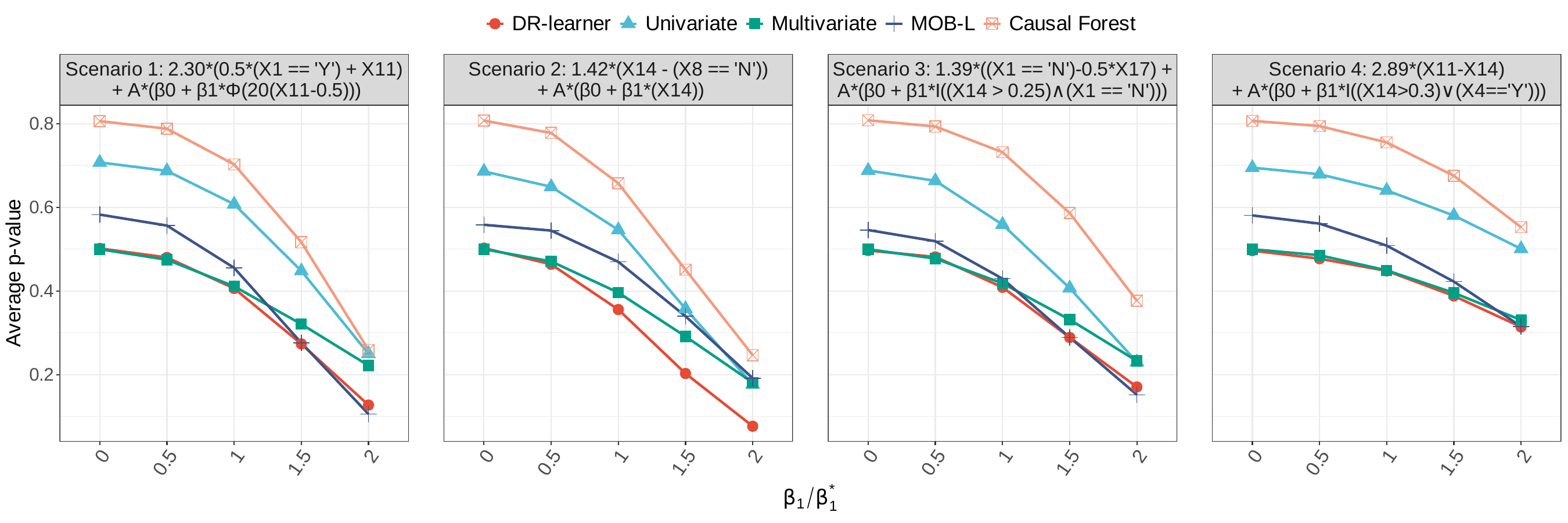}
  \end{center}
      \vspace{-0.5cm}
  \caption{\textbf{Comparison of our proposed DR-learner with competing methods regarding Objective 1(ii)}. Data are simulated under various degrees of treatment effect heterogeneity (x-axis). We report the average p-values (across 500 runs) and when there is treatment effect heterogeneity (i.e., $\beta_1 > 0$), the lower the p-value, the more powerful the method.}
   \label{fig:competing_methods_obj_1ii}
       \vspace{-0.5cm}
\end{figure}

Regarding Objective 1(i), as illustrated in Figure \ref{fig:competing_methods_obj_1i}, only the Multivariate and DR-learner methods achieve a uniform distribution of p-values under the null hypothesis of no TEH. The other methods tend to skew towards higher p-values, making them more conservative. Furthermore, to investigate whether any of the methods lead to inflated false positive rates, we present the results in Table \ref{tbl:competing_methods_obj_1i}. This table shows the estimated false positive rate when the nominal level is $\alpha=0.10$. The numbers represent the proportion of times that the $p$-value is less than or equal to 0.10. As we can see, all the methods provide estimates lower than the nominal level, indicating that they do not lead to an inflation of type-I error. For Objective 1(ii), Figure \ref{fig:competing_methods_obj_1ii} demonstrates that our DR-learner consistently performs excellently, ranking in the top two methods across  scenarios and degrees of TEH. In contrast, the Causal Forest and Univariate methods do not achieve comparable performance.

\begin{table}[ht]
\centering
\caption{\textbf{Checking whether the methods lead to inflation of type-I error}. For the data presented in Figure \ref{fig:competing_methods_obj_1i} we estimate the false positive rates for nominal level of $\alpha = 0.10.$}
  \label{tbl:competing_methods_obj_1i}
\begin{tabular}{lrrrrr}
  \hline
 & DR-learner & Univariate & Multivariate & MOB-L & Causal Forest \\ 
  \hline
Scenario 1:
$2.30*(0.5(X1 == 'Y') + X11) + A\beta_0$ & 0.08 & 0.10 & 0.09 & 0.09 & 0.04 \\ 
  Scenario 2:
$1.42*(X14 - (X8 == 'N')) + A\beta_0$  & 0.09 & 0.09 & 0.09 & 0.08 & 0.04 \\ 
  Scenario 3:
$1.39*((X1 == 'N')-0.5X17) + A\beta_0$  & 0.08 & 0.07 & 0.09 & 0.08 & 0.05 \\ 
  Scenario 4:
$2.89*(X11-X14) + A\beta_0$  & 0.08 & 0.09 & 0.09 & 0.08 & 0.04 \\ 
   \hline
\end{tabular}
\end{table}

\subsubsection{Results Objective 2}
\begin{figure}[!h]
  \begin{center}
    \includegraphics[width=0.95\textwidth]{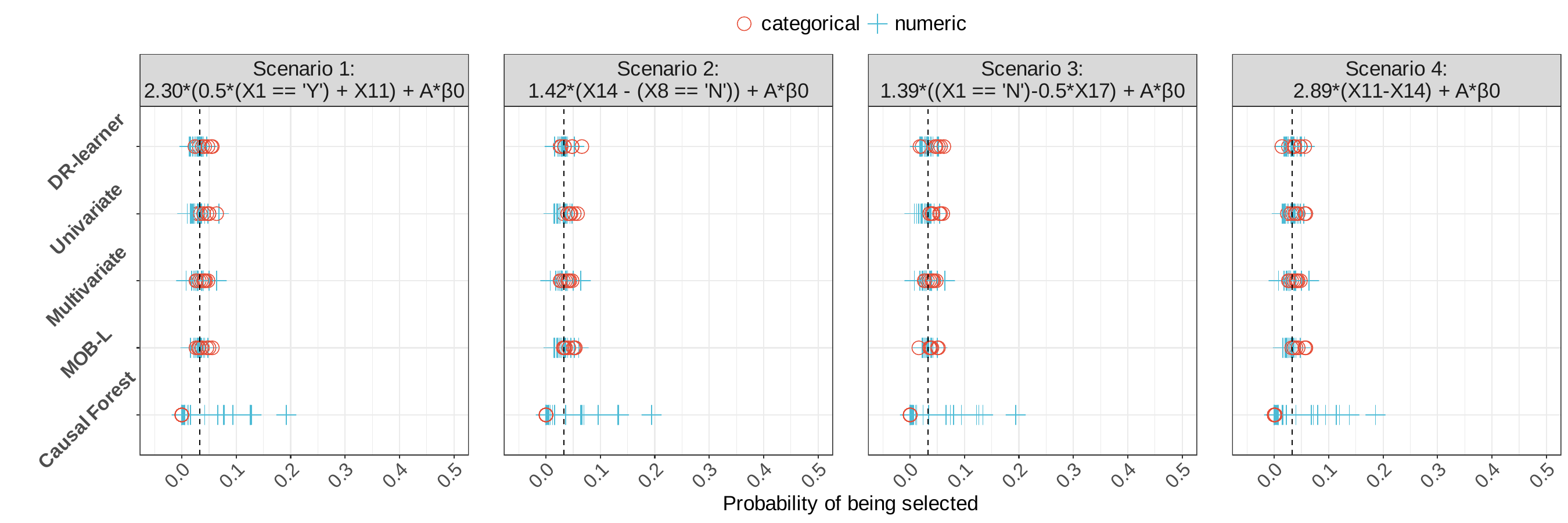}
  \end{center}
    \vspace{-0.5cm}
  \caption{\textbf{Comparison of our proposed DR-learner with competing methods regarding Objective 2(i)}. Data are simulated under the condition of no treatment effect heterogeneity (i.e., $\beta_1 = 0$). We report the average probability (across 500 runs) that each biomarker is selected as the most important predictive biomarker. Since there is no treatment effect heterogeneity, for a method to be unbiased all biomarkers should have probability equal to $1/30 \approx 0.03,$ dashed vertical line. The Causal Forest almost never selected any categorical variables as the most important, which is why there are no red circles are in $0$.}
  \label{fig:competing_methods_obj_2i}
\end{figure}

\begin{figure}[!h]
  \begin{center}
    \includegraphics[width=0.99\textwidth]{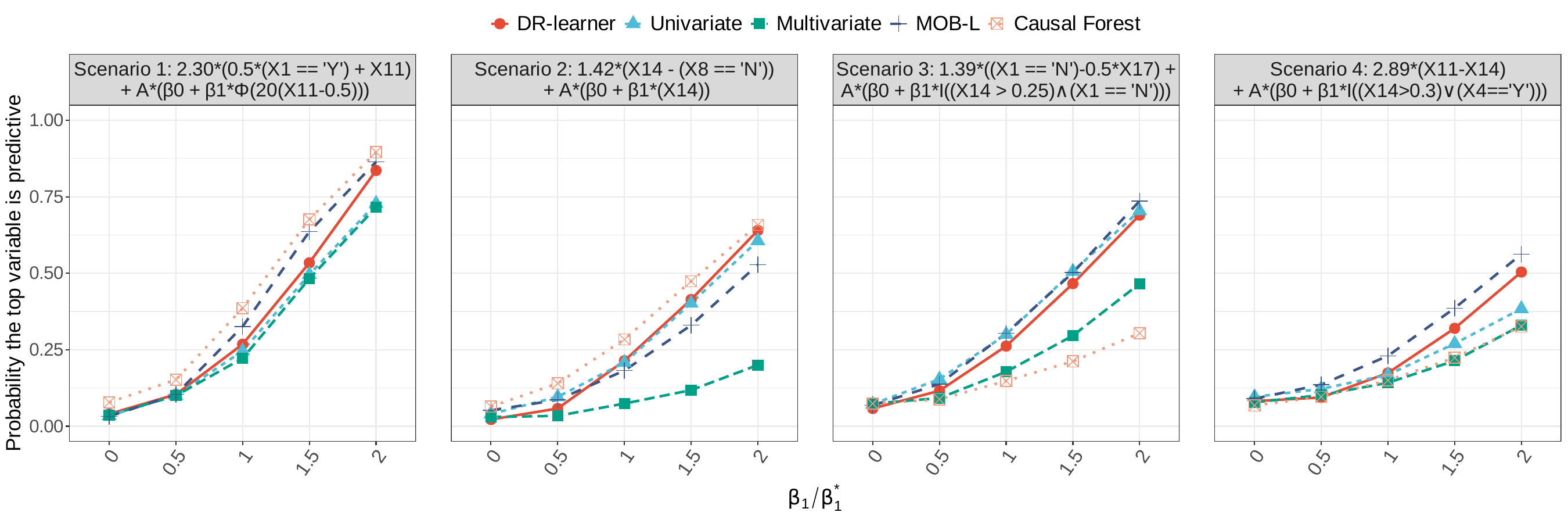}
  \end{center}
  \caption{\textbf{Comparison of our proposed DR-learner with competing methods regarding Objective 2(ii)}. Data are simulated under various degrees of treatment effect heterogeneity (x-axis). We report the average (across 500 runs) probability that the top selected biomarker is truly predictive, and the higher this probability are the better the performance.}
   \label{fig:competing_methods_obj_2ii}
\end{figure}

For Objective 2(i), as illustrated in Figure \ref{fig:competing_methods_obj_2i}, all methods are unbiased except for the Causal Forest, which shows a clear preference for continuous biomarkers. For instance, in Scenarios 1-3, across 500 runs without TEH, it never selects any of the categorical variables (red circle). Regarding Objective 2(ii), as shown in \ref{fig:competing_methods_obj_2ii}, our DR-learner consistently achieves comparable performance, always ranking among the top 2-3 methods. In contrast, other methods show inconsistent performance across different scenarios. For example, the Causal Forest performs well in Scenarios 1 and 2, where the effect modifiers are continuous, but fails to achieve comparable performance in Scenarios 3 and 4, which include categorical effect modifiers. 

\subsubsection{Results Objective 3}
\begin{figure}[!h]
  \begin{center}
    \includegraphics[width=0.99\textwidth]{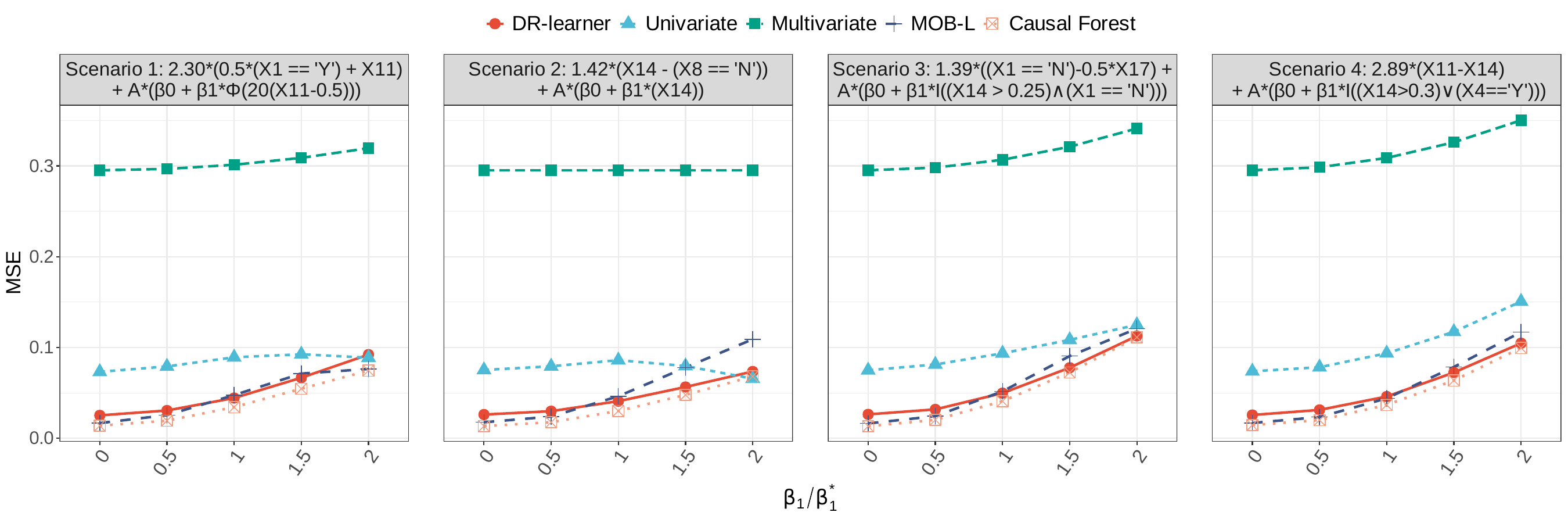}
  \end{center}
    \caption{\textbf{Comparison of our proposed DR-learner with competing methods regarding Objective 3 with respect of the MSE}. Data are simulated under various degrees of treatment effect heterogeneity (x-axis). We report the MSE (across 500 runs) for estimating CATE, and the lower this error is the better the performance of the method.}
  \label{fig:competing_methods_obj_3_mse}
\end{figure}

\begin{figure}[!h]
  \begin{center}
    \includegraphics[width=0.99\textwidth]{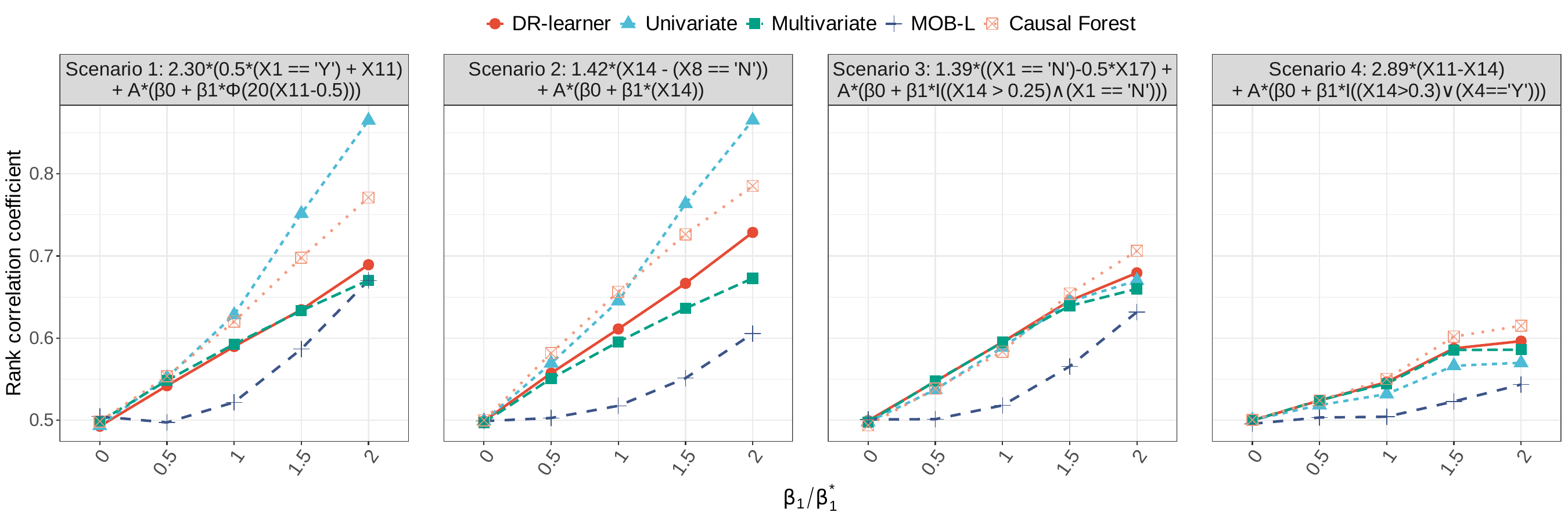}
  \end{center}
    \caption{
    { \textbf{Comparison of our proposed DR-learner with competing methods regarding Objective 3 with respect of Somers' D rank correlation coefficient}. Data are simulated under various degrees of treatment effect heterogeneity (x-axis). We report the Somers' D rank correlation coefficient(across 500 runs) for estimating CATE, and the larger this index is the better the performance of the method.}}
  \label{fig:competing_methods_obj_3_rank_correlation}
\end{figure}

{ Finally, for Objective 3, the performance of the competing methods is illustrated in Figures \ref{fig:competing_methods_obj_3_mse} and \ref{fig:competing_methods_obj_3_rank_correlation}, which present results based on MSE and rank correlation coefficient, respectively. Across both evaluation metrics, the DR-learner demonstrates good and consistent performance, closely approaching that of the Causal Forest, which remains the top-performing method throughout.}

\subsubsection{Performance summary of DR-learner and alternatives}
To provide a high-level summary of the results, we have created Table \ref{table:competing_methods}. This table shows that the DR-learner is the only method that consistently achieves excellent or competitive performance. In contrast, the other methods  fail to achieve competitive performance in at least one objective. Notably, the Causal Forest method performs well only in Objective 3, which makes sense since the method is designed to focus on estimating CATE. The MOB-L method shows mixed results.
\begin{table}[!h]
\caption{\textbf{Comparison of methods across objectives.} The symbols used are as follows:\\
\circledcheck~indicates excellent performance, consistently ranking among the top-2 methods in all scenarios across all levels of TEH.\\
\circledO~indicates competitive performance, consistently ranking among the top-3 methods in all scenarios across all levels of TEH.\\
\circledX~indicates that the method failed to achieve competitive performance.}
\label{table:competing_methods}
\centering
%\Large
\setlength{\extrarowheight}{3pt}
\begin{tabular}{lccccc}
\toprule
\textbf{Method} & \textbf{Objective 1(i)} & \textbf{Objective 1(ii)} & \textbf{Objective 2(i)} & \textbf{Objective 2(ii)} & \textbf{Objective 3} \\
\midrule
Univariate & \circledX & \circledX & \circledcheck & \circledX & \circledX \\
Multivariate & \circledcheck &  \circledO & \circledcheck &\circledX & \circledX \\
MOB-L & \circledO & \circledO & \circledcheck & \circledX & \circledO \\
Causal Forest & \circledX & \circledX & \circledX & \circledX & \circledcheck\\
DR-learner & \circledcheck & \circledcheck & \circledcheck & \circledO &  \circledO \\
\bottomrule
\end{tabular} 
\end{table}

\section{Clinical trial example}
\label{sec:clinical_trial}
Psoriatic arthritis (PsA) is a chronic inflammatory condition that impacts the joints, entheses, and skin, leading to impaired physical function.\cite{future2} PsA significantly affects the quality of life for patients, manifesting symptoms in various parts of the body, including the joints, spine, skin, and nails. Early detection and treatment are crucial to prevent permanent joint damage. The primary endpoint for assessing treatment improvement is the binary score ACR50, developed by the American College of Rheumatology.\cite{felson1993american} This composite score indicates a 50\% improvement in the number of tender and swollen joints, as well as a 50\% improvement in three out of five criteria: patient global assessment, physician global assessment, functional ability measure (typically the Health Assessment Questionnaire, HAQ), visual analog pain scale, and high sensitivity C-reactive protein (hsCRP) or erythrocyte sedimentation rate. If these improvements are met, the score is $Y=1$; otherwise, it is $Y=0$.

Cosentyx (secukinumab) is approved for treating adult patients with active psoriatic arthritis and has been evaluated in several clinical trials. In this study, we analyze data from five Phase III trials: FUTURE 1,\cite{future1} FUTURE 2,\cite{future2} FUTURE 3,\cite{future3} FUTURE 4,\cite{future4} and FUTURE 5.\cite{future5}  Our research objective is to identify markers that indicate treatment heterogeneity for ACR50 at week 16 when comparing Placebo ($T=0$) versus Cosentyx ($T=1$), regardless of dosage. The estimand of interest is the risk difference, i.e. $ \text{Pr}(Y=1|T=1) - \text{Pr}(Y=1|T=0).$ 

{ Since CATE is defined as the difference in conditional expectations of the outcome - given specific covariate values - it is applicable not only to continuous outcomes but also to binary ones, when the estimand of interest is the risk difference. %A key advantage of CATE is that it is model-agnostic: its definition does not rely on any specific statistical model. 
From a public health perspective, the absolute scale - such as the risk difference - is often considered the most relevant for measuring treatment effects. This is because differences in expectations (e.g., probabilities in the case of binary outcomes) can be directly interpreted in terms of the number of individuals in a population who would experience or avoid a particular event due to the treatment.\cite{vanderweele2014tutorial}}

There are three prior studies that analyze FUTURE trials with objectives similar to ours. Sechidis et al.\cite{Sechidis2021} used the knockoff methodology to identify clinical variables that act as effect modifiers. They did a pooled analysis (all trials except FUTURE 1) while controlling the false discovery rate at 20\%. They identified eight effect modifiers: C-reactive protein, age, polyarticular arthritis, asymmetric peripheral arthritis, psoriasis nail subset, sex, fatigue score, and body surface area (see Figure 5 in the aforementioned work\cite{Sechidis2021}). Bornkamp et al.\cite{bornkamp2023predicting} used the same data for a subgroup analysis challenge. The teams provided various solutions, and in terms of the effect modifiers, the main insights were very similar to those in the previous work. For example, the variables most commonly used across the teams to define subgroups were C-reactive protein, age, and fatigue score (see Figure 2B in the aforementioned work\cite{bornkamp2023predicting}). Finally, Cardner et al.\cite{Cardner2023} analyzed proteomics data from serum samples. They trained a stability selection model on all trials except FUTURE 2 and identified beta-defensin 2 (BD-2) in serum as a promising effect modifier; higher baseline levels of BD-2 are robustly linked to better clinical outcomes with Secukinumab, but not with placebo.

For us, it would be insightful to assess TEH in these data using our DR-learner approach, and especially explore whether it identifies similar effect modifiers to those reported in the literature. To achieve this, we compiled a dataset from all trials, including 1937 patients. After preprocessing—such as removing covariates with more than 20\% missing data—we used multiple imputation to address missing values in the remaining covariates. The imputed values were then aggregated by taking the median for numeric variables and the mode for categorical ones, resulting in a single completed dataset with 70 covariates of mixed type. These included a variety of variables, such as demographic information (e.g., age, sex, BMI), medical history (e.g., time since first PsA diagnosis, presence of psoriasis, presence of polyarticular arthritis), treatment and medication details (e.g., corticosteroid use, TNF alpha inhibitors), as well as lab values, clinical measurements, quality of life assessments, efficacy measurements, and the proteomic marker BD-2.

\begin{figure}[!h]
  \begin{center}
    \includegraphics[width=0.55\textwidth]{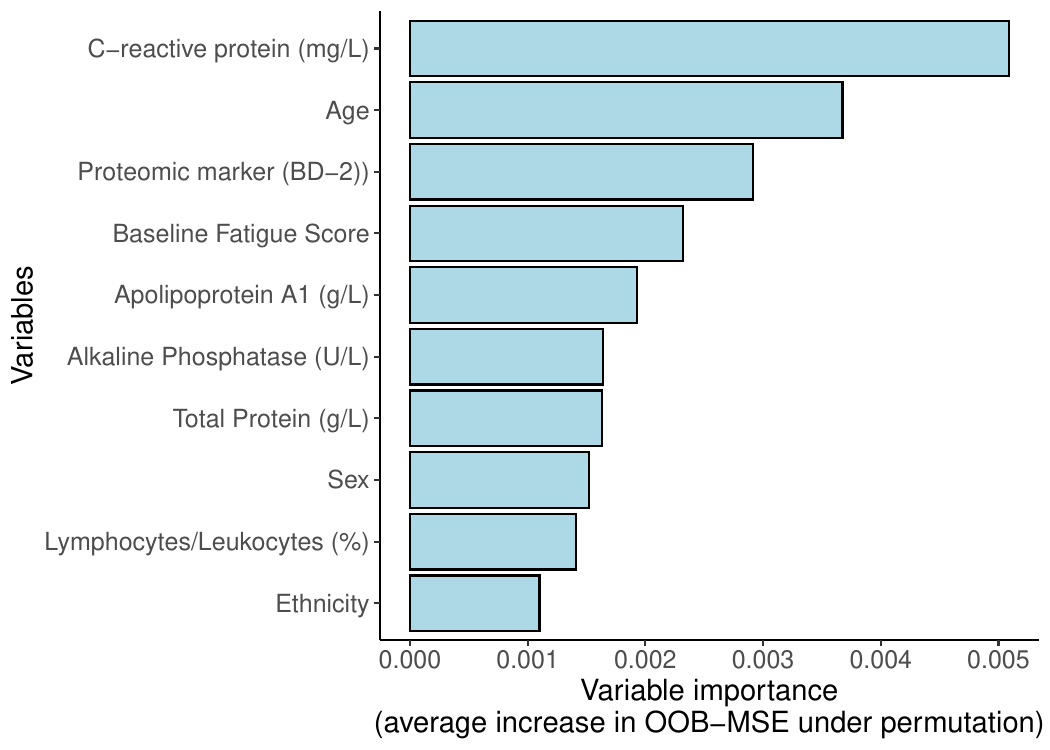}
  \end{center}
  \caption{\textbf{Variable importance ranking returned from the DR-learner.} This figure shows the top 10 variables ranked by their importance, emphasizing their roles as effect modifiers. Higher-ranked variables indicate stronger modification effects.}
  \label{fig:clinical_trial_importance_scores}
\end{figure}

{ To assess TEH we applied the proposed DR-learner approach, as described in Section \ref{sec:sim:competing_methods}.}  Regarding \textbf{Objective 1}, the overall assessment provides a very low p-value ($2 \times 10^{-5}$), which is reassuring since we already have evidence from previously published works that various markers modify treatment effects. The more interesting insights come when we move to \textbf{Objective 2} and the identification of possible effect modifiers. Figure \ref{fig:clinical_trial_importance_scores} shows the ranking of the top five variables returned by the DR-learner. The top two variables, C-reactive protein (CRPSI) and Age, are also the top discoveries in two previous works \cite{Sechidis2021,bornkamp2023predicting}, while the Baseline Fatigue Score (FACITSO) was also among the top variables in the previous works. Additionally, the DR-learner ranked the proteomic marker BD-2 in the third position, aligning with the previous work that identified BD-2 as a potential effect modifier.\cite{Cardner2023} Overall, the DR-learner effectively identified effect modifiers that have been reported in the literature. 

\begin{figure}[!h]
  \begin{center}
    \includegraphics[width=0.49\textwidth]{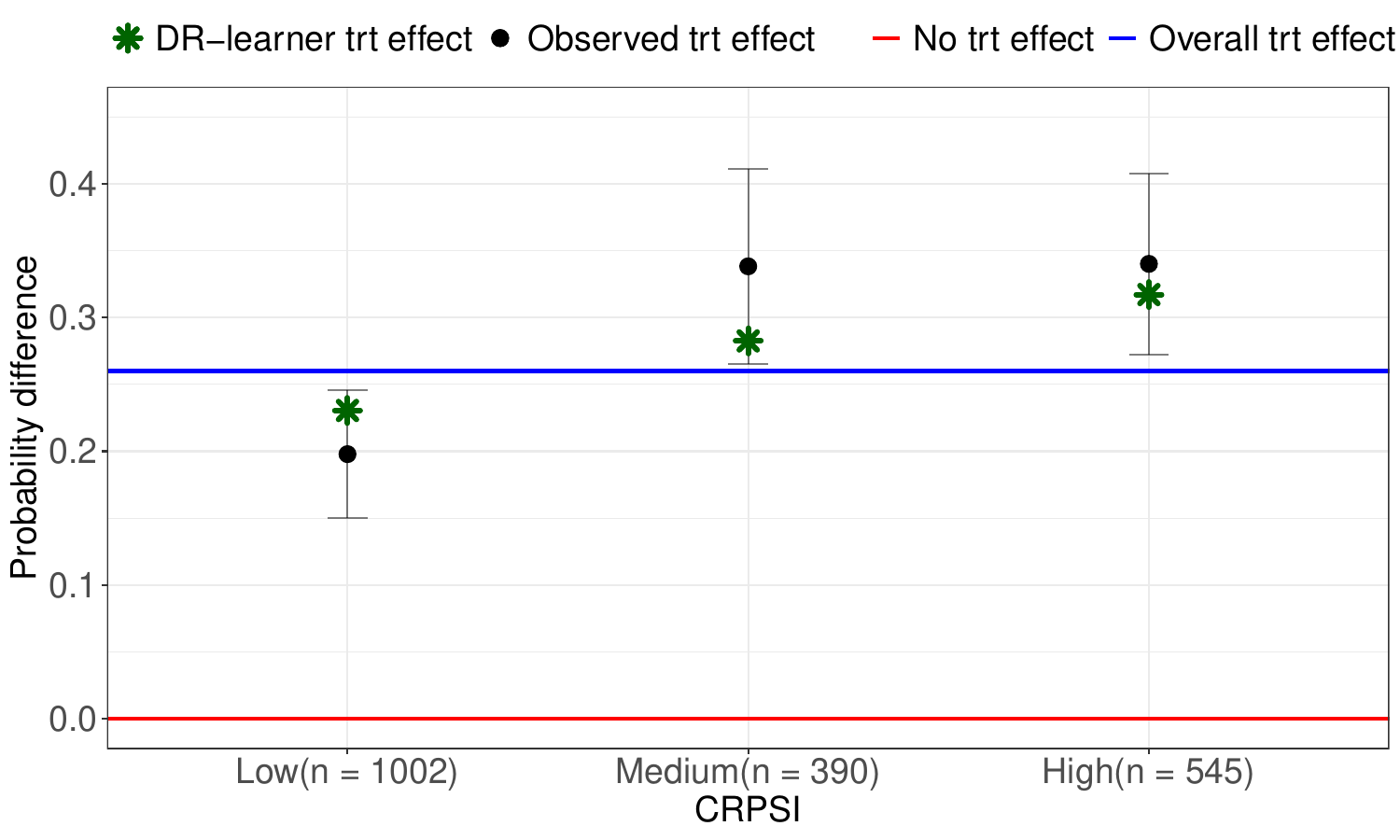}
    \hfill  
    \includegraphics[width=0.49\textwidth]{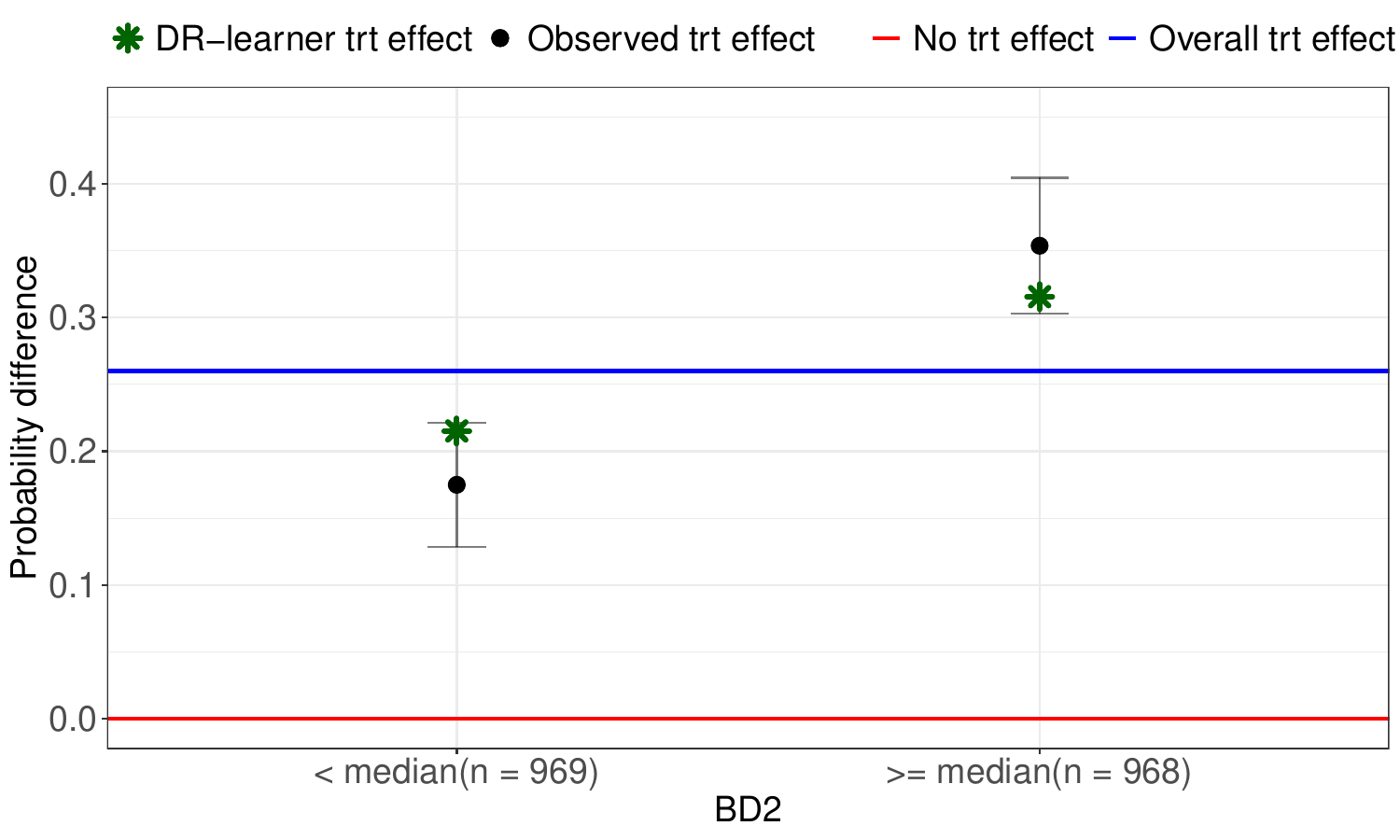}
  \end{center}
  \caption{\textbf{Displays that capture how treatment effect changes with CRPSI and the proteomic marker (BD-2)}. We provide (i) the estimated treatment effect from the DR-learner (green symbol/line), (ii) the observed estimates (unadjusted) of the treatment effect (black symbol/line) with confidence intervals, (iii) for reference, the line of no treatment effect (red line), and (iv) for reference, the line of the overall treatment effect (ATE) (blue line).}
  \label{fig:clinical_trial_importance_scores_CRPSI_BD2}
\end{figure}

Finally, it will be interesting to explore how the identified variables change the treatment effect. Figure \ref{fig:clinical_trial_importance_scores_CRPSI_BD2} provides these visualizations for CRPSI and BD-2. For the first variable, we use the same categories as Sechidis et al.\cite{Sechidis2021} and observe very similar results, i.e., the higher the CRPSI baseline value, the larger the treatment effect. An interesting caveat is that the estimated treatment effect from the DR-learner, which provides an adjusted way for estimation, shrinks the treatment effects in the different subgroups towards the overall effect. We observe a similar trend for the continuous marker BD-2, and again our results are in line with Cardner et al.\cite{Cardner2023}; patients with above-median levels of BD-2 at baseline had a higher treatment effect.

\section{Conclusions}
\label{sec:conclusions}
In this work we demonstrated how the DR-learner could be effectively utilized within the WATCH workflow to provide a comprehensive framework for assessing treatment effect heterogeneity. The proposed methods addressed three key objectives: performing global tests for homogeneity, ranking covariates based on their influence on effect modification, and estimating individualized treatment effects.

By employing a semi-parametric and robust implementation of the DR-learner with SuperLearner, we derived pseudo-outcomes that can be seen as patient level causal contrast in expectation. These pseudo-outcomes were used to conduct a global assessment against homogeneity using conditional inference permutation test, addressing the first objective. For the second objective, we formulated a regression problem and fitted random forests built with conditional inference trees, deriving permutation-based importance scores that were unbiased towards variable types and captured the strength of baseline covariates in modifying the treatment effect. Finally, the DR-learner was used to estimate CATE, capturing individualized treatment effects.  Our simulation study evaluated the operating characteristics of the suggested methods across various objectives and compared them with several competing methods. While we primarily considered RCT scenarios, it is important to emphasize that the suggested methodology is also applicable to observational settings.

Additionally, we applied our workflow to a large pool study of trials in Psoriatic Arthritis, demonstrating its utility in assessing heterogeneity. Through this case study we illustrated also another possible usage of our workflow; it can be used to assess the effect modification strength of omics biomarkers in the context of clinical variables.  The importance of this type of approaches has been emphasized in other contexts. For example, Boulesteix and Sauerbrei \cite{Boulesteix2011} emphasize the need to evaluate the added predictive value of molecular signatures alongside existing clinical predictors for predicting patient outcomes in clinical settings. Having this type of evaluation in the context of discovering biomarkers that act as effect modifiers is crucial.

{ Future research should explore the application of our methodologies to a broader range of clinical endpoints and treatment effect measures. Our current work focuses on estimating the conditional treatment effect, specifically CATE, which is defined as the difference in conditional expectations of the outcome given covariates. While this can be seen as a natural measure for continuous endpoints, and also for binary when the interest is in risk difference,  it is not straightforward how conditional treatment effect measures can be extended to other endpoints, for example time-to-event, count or binary when the interest is the odds ratios.} To the best of our knowledge, there are no works that directly use the DR-learner to derive pseudo-outcomes for time-to-event endpoints that could be directly integrated into the workflow we presented in this paper. One promising direction for handling time-to-event endpoints is the recent work of Pryce et al. \cite{pryce2025causal}, which extends the DR-learner to the missing-at-random (MAR) outcome setting. This method could potentially be applied to time-to-event data by treating censored observations as missing, although this may discard some information. Furthermore, there is a wealth of recent research\cite{bo2023meta, li2024doublyrobusttargetedestimation, gao2024doubly, ziersen2024variableimportancemeasuresheterogeneous} that can serve as a foundation for exploring how a WATCH-type workflow can be applied to analyze survival data in line with our three objectives.

{ A different approach to implementing the WATCH workflow for alternative effect measures and endpoints involves relying on a parametric regression model. This type of solution is presented in a companion paper, \cite{chen2025comparingmethodsassesstreatment} which compares methods to assess treatment effect heterogeneity in general parametric regression models.}

Another area worth further exploration is using alternative ways for deriving effect modifiers in objective 2. For example, instead of our approach that relies on marginal permutation, we can leverage conditional randomization inference methods,\cite{candes2018panning, Floodgate2020, liu2022fast,zimmermann2024all} or other alternatives.\cite{ hama2023modelfreevariableimportance, verdinelli2024decorrelated} 
{ If there is interest in selecting a subset of effect modifiers rather than providing a full ranking, several methods have been proposed in recent years that enable such selection while controlling the risk of Type I error.\cite{Sechidis2021, paillard2024} Finally, while our focus has been on evaluating DR learner, we acknowledge that simpler modeling strategies may offer practical alternatives in certain settings. Future work could explore a broader range of baseline methods, including those based on standard regression and machine learning approaches, to further contextualize the performance of the different estimators.}
%Furthermore, since our objective is to derive rankings that capture  the strength of effect modification, we use methods that have been suggested recently to answer more directly this question recent years, various methods have been suggested that directly address this question, such as the works of Hines et al.\cite{hines2022variable} and Paillard et al.\cite{paillard2024} Furthermore, instead of the permutation importance measure we are currently using, we can consider other alternative measures that have been recently suggested.

\section*{Acknowledgments}
The authors would like to thank Alain Marti for his feedback on the manuscript and Torsten Hothorn, Jiarui Lu and Oliver Dukes for the helpful discussions during the course of this work. We also thank the EFSPI Treatment Effect Heterogeneity Special Interest Group (SIG) for their many discussions and feedback.

\section*{Dava availability statement}
Due to confidentiality and privacy concerns, the patient level data from the application section cannot be shared publicly. The code that implements the DR-learner will be available at: https://github.com/Novartis/WATCH/.

\appendix
\section{Prove the equivalence of two expressions of the DR-learner}
\label{app:proof} 
Kennedy\cite{kennedy2023towards} provide an expression that can be seen as a correction of the outcome models:
\begin{align}
\widehat{\psi}_{\text{DR:1}} (\mathbf{x}_i,y_i) &= \frac{a_i - \widehat{\pi}(\mathbf{x}_i)}{\widehat{\pi}(\mathbf{x}_i)\left(1-\widehat{\pi}(\mathbf{x}_i)\right)} \left(y_i - \widehat{\mu}_{a_i} (\mathbf{x}_i) \right) + \widehat{\mu}_1 (\mathbf{x}_i) - \widehat{\mu}_0 (\mathbf{x}_i) \notag 
\end{align}
Curth and Van der Schaar\cite{curth2021nonparametric} provide an expression that can be seen as a correction of the IPWE estimator:
\begin{align}
 \widehat{\psi}_{\text{DR:2}} (\mathbf{x}_i,y_i) &=\frac{a_i - \widehat{\pi}(\mathbf{x}_i)}{\widehat{\pi}(\mathbf{x}_i)\left(1-\widehat{\pi}(\mathbf{x}_i)\right)} y_i  +
\left( \left(1 - \frac{a_i}{\widehat{\pi}(\mathbf{x}_i)} 
\right)\widehat{\mu}_1 (\mathbf{x}_i) - \left(1 - \frac{1-a_i}{1-\widehat{\pi}(\mathbf{x}_i)} 
\right)\widehat{\mu}_0(\mathbf{x}_i) \right) \notag 
\end{align}

We will show that these two expression are equivalent. Let's start from the first expression:
\begin{align}
\widehat{\psi}_{\text{DR:1}} (\mathbf{x}_i,y_i) &= \frac{a_i - \widehat{\pi}(\mathbf{x}_i)}{\widehat{\pi}(\mathbf{x}_i)\left(1-\widehat{\pi}(\mathbf{x}_i)\right)} \left(y_i - \widehat{\mu}_{a_i} (\mathbf{x}_i) \right) + \widehat{\mu}_1 (\mathbf{x}_i) - \widehat{\mu}_0 (\mathbf{x}_i) \notag \\
\widehat{\psi}_{\text{DR:1}} (\mathbf{x}_i,y_i) &= \frac{a_i - \widehat{\pi}(\mathbf{x}_i)}{\widehat{\pi}(\mathbf{x}_i)\left(1-\widehat{\pi}(\mathbf{x}_i)\right)} \left(y_i - a_i \widehat{\mu}_1 (\mathbf{x}_i) -(1-a_i)\widehat{\mu}_0 (\mathbf{x}_i) \right) + \widehat{\mu}_1 (\mathbf{x}_i) - \widehat{\mu}_0 (\mathbf{x}_i) \notag \\
\widehat{\psi}_{\text{DR:1}} (\mathbf{x}_i,y_i) &= \left( \frac{a_i}{\widehat{\pi}(\mathbf{x}_i)} - \frac{1-a_i}{1-\widehat{\pi}(\mathbf{x}_i)} \right) \left(y_i - a_i \widehat{\mu}_1 (\mathbf{x}_i) -(1-a_i)\widehat{\mu}_0 (\mathbf{x}_i) \right) + \widehat{\mu}_1 (\mathbf{x}_i) - \widehat{\mu}_0 (\mathbf{x}_i) \notag 
\end{align}

Expanding this, we get:
\begin{align}
\widehat{\psi}_{\text{DR:1}} (\mathbf{x}_i,y_i) &= 
\frac{a_i}{\widehat{\pi}(\mathbf{x}_i)}y_i -  \frac{a_i^2}{\widehat{\pi}(\mathbf{x}_i)}\widehat{\mu}_1 (\mathbf{x}_i) -  \frac{a_i(1-a_i)}{\widehat{\pi}(\mathbf{x}_i)}\widehat{\mu}_0 (\mathbf{x}_i) 
-\frac{1-a_i}{1-\widehat{\pi}(\mathbf{x}_i)} y_i + \frac{(1-a_i)a_i }{1-\widehat{\pi}(\mathbf{x}_i)}\widehat{\mu}_1 (\mathbf{x}_i) + \frac{(1-a_i)^2}{1-\widehat{\pi}(\mathbf{x}_i)}\widehat{\mu}_0 (\mathbf{x}_i) 
+ \widehat{\mu}_1 (\mathbf{x}_i) - \widehat{\mu}_0 (\mathbf{x}_i) \notag 
\end{align}
Since \( a_i \) takes values 0 and 1, \( a_i^2 = a_i \),  \( (1-a_i)^2 = 1-a_i \) and \( a_i(1-a_i) = 0 \). Therefore, the expression simplifies to:
\begin{align}
\widehat{\psi}_{\text{DR:1}} (\mathbf{x}_i,y_i) &= 
\frac{a_i}{\widehat{\pi}(\mathbf{x}_i)}y_i -  \frac{a_i}{\widehat{\pi}(\mathbf{x}_i)}\widehat{\mu}_1 (\mathbf{x}_i)
-\frac{1-a_i}{1-\widehat{\pi}(\mathbf{x}_i)} y_i + \frac{(1-a_i)}{1-\widehat{\pi}(\mathbf{x}_i)}\widehat{\mu}_0 (\mathbf{x}_i) 
+ \widehat{\mu}_1 (\mathbf{x}_i) - \widehat{\mu}_0 (\mathbf{x}_i) \notag \\
\widehat{\psi}_{\text{DR:1}} (\mathbf{x}_i,y_i) &= 
\left( \frac{a_i}{\widehat{\pi}(\mathbf{x}_i)} - \frac{1-a_i}{1-\widehat{\pi}(\mathbf{x}_i)}\right)y_i + \left(1 -\frac{a_i}{\widehat{\pi}(\mathbf{x}_i)}\right) \widehat{\mu}_1 (\mathbf{x}_i) - \left(1 - \frac{1-a_i}{1-\widehat{\pi}(\mathbf{x}_i)} \right) \widehat{\mu}_0 (\mathbf{x}_i) \notag\\
\widehat{\psi}_{\text{DR:1}} (\mathbf{x}_i,y_i) &= 
 \frac{a_i - \widehat{\pi}(\mathbf{x}_i)}{\widehat{\pi}(\mathbf{x}_i)\left(1-\widehat{\pi}(\mathbf{x}_i)\right)} y_i + \left(1 -\frac{a_i}{\widehat{\pi}(\mathbf{x}_i)}\right) \widehat{\mu}_1 (\mathbf{x}_i) - \left(1 - \frac{1-a_i}{1-\widehat{\pi}(\mathbf{x}_i)} \right) \widehat{\mu}_0 (\mathbf{x}_i) \notag\\
\widehat{\psi}_{\text{DR:1}} (\mathbf{x}_i,y_i) &= \widehat{\psi}_{\text{DR:2}} (\mathbf{x}_i, y_i) \notag
\end{align}

{
\section{Illustrative Simulation Example}
To provide more intuition about the suggested methods, we conducted a toy simulation based on Scenario 2 with $\beta = 2$, described in Table \ref{tab:sim_models}. In this scenario the heterogeneity of treatment effect is driven by a single predictive covariate ($X_{14}$). This variable also acts as a prognostic factor, alongside $X_8$, which is prognostic only.

Furthermore, to illustrate the impact of various parameters in the estimation of CATE, we simulate 3 different settings.
\begin{itemize}
    \item \textbf{Setting 1:} 500 patients, using all 30 baseline covariates.
    \item \textbf{Setting 2:} 1000 patients, using all 30 baseline covariates.
    \item \textbf{Setting 3:} 500 patients, using only covariates $X_8$ and $X_{14}$ (prognostic and/or predictive).
\end{itemize}

The motivation behind these scenarios is to evaluate how the proposed methods perform under different data conditions. Setting 2 represents a larger trial with two times the sample size of Setting 1, which is expected to enhance the precision of CATE estimates due to increased statistical power. Setting 3, on the other hand, by modelling only the true prognostic and predictive covariates ($X_8$ and $X_{14}$), reduces model complexity and eliminating noise from irrelevant features. Both settings are therefore considered more favorable than Setting 1, albeit for different reasons: Setting 2 benefits from the larger sample size, while Setting 3 by only leveraging variables that are aligned with the true data-generating process. For each setting, we estimated pseudo-outcomes ($\widehat{\phi}$) and CATEs ($\widehat{\tau}$) using the DR-learner presented in Algorithm \ref{alg:DR_learner}. Table~\ref{tab:simulation_outputs} presents a sample of the simulated data and model outputs.
\begin{table}[h!]
    \centering
    \caption{Sample of simulated data and model outputs for each setting.}
    \label{tab:simulation_outputs}
    \begin{tabular}{c|cccc|c||cc||cc||cc}
        \toprule
         Patient &
        \multicolumn{4}{c|}{\textbf{Input Variables}} & 
        \textbf{True CATE} &
        \multicolumn{2}{c||}{\textbf{Setting 1}} & 
        \multicolumn{2}{c||}{\textbf{Setting 2}} & 
        \multicolumn{2}{c}{\textbf{Setting 3}} \\
        \cmidrule(r){2-5} \cmidrule(lr){7-8} \cmidrule(lr){9-10} \cmidrule(l){11-12}
        ID &$X_8$ & $X_{14}$ & Trt & $Y$ &$\tau$  & $\widehat{\tau}_1$ & $\widehat{\phi}_1$ & $\widehat{\tau}_2$ & $\widehat{\phi}_2$ & $\widehat{\tau}_3$ & $\widehat{\phi}_3$ \\
        \midrule
        % Example row
1 & N & 0.350 & 1.476 & 0 & 0.219 & 0.404 & -4.231 &  0.373 & -4.914 & 0.283 & -4.349 \\
2 & N & 0.470 & -1.081 & 0 & 0.483 & 0.388 &  1.435 & 0.431 & 1.250 & 0.406 &  1.318 \\
3 & N & 0.281 & 1.489 & 0 & 0.067 & 0.146 & -4.466 & 0.017 & -4.626 & 0.237 & -4.802 \\
4 & N & 0.395 & -0.383 & 0 & 0.317 & 0.389 & -0.533 & 0.365 & -0.292 & 0.208 & -0.454 \\
5 & N & 0.258 & -0.427 & 1 & 0.016 & 0.194 &  0.805 &  0.192 & 0.858 & 0.164 &  0.998 \\
6 & N & 0.448 & -1.297 & 0 & 0.435 & 0.216 &  1.619 &  0.421 & 1.406 & 0.386 &  1.495 \\
7 & Y & 0.078 & 0.560 & 0 & -0.381 & -0.039 & -1.096 & 0.150 & -1.285 & -0.205 & -1.164 \\
8 & N & 0.600 & 0.046 & 0 & 0.770 & 0.369 & -0.399 & 0.296 & -0.385 & 0.593 & -0.268 \\
9 & Y & 0.454 & 0.140 & 0 & 0.448  & 0.192 &  0.914 & 0.339 & 1.057 & 0.119 &  1.425 \\
10 & N & 0.380 & -2.223 & 1 & 0.284 & -0.008 & -2.866 & 0.209 & -2.981 & 0.198 & -3.171 \\
        \bottomrule
        \multicolumn{6}{c||}{\textbf{Overall performance measures}} & \multicolumn{2}{c||}{}& \multicolumn{2}{c||}{}& \multicolumn{2}{c}{}\\
        \multicolumn{6}{c||}{\textbf{Objective 1}: p-value from the overall test against homogeneity} & \multicolumn{2}{c||}{0.17}& \multicolumn{2}{c||}{$<$0.0001}   & \multicolumn{2}{c}{0.01}\\
        \multicolumn{6}{c||}{\textbf{Objective 2}: Top selected biomarker as predictive} & \multicolumn{2}{c||}{$X_{14}$}& \multicolumn{2}{c||}{$X_{14}$}& \multicolumn{2}{c}{$X_{14}$}\\
         \multicolumn{6}{c||}{\textbf{Objective 3}: Evaluate estimated CATEs using $\text{MSE}(\tau, \widehat{\tau})$} & \multicolumn{2}{c||}{0.094}& \multicolumn{2}{c||}{0.067}& \multicolumn{2}{c}{0.047}\\
        \multicolumn{6}{c||}{\textbf{Objective 3}: Evaluate estimated CATEs using } & \multicolumn{2}{c||}{0.661}& \multicolumn{2}{c||}{0.771}& \multicolumn{2}{c}{0.881}\\
        \multicolumn{6}{c||}{Somers' D rank corr. coefficient $D(\tau, \widehat{\tau})$} & &&\\

    \end{tabular}
\end{table}

Furthermore, Table \ref{tab:sim_models} presents various performance metrics across the three simulation settings. For the overall test of treatment effect heterogeneity, Setting 1 yields the highest p-value ($0.17$), suggesting weaker evidence for heterogeneity, while Settings 2 and 3 produce substantially lower p-values, indicating stronger signals. Notably, in all three scenarios, the top-ranked biomarker is the true predictive covariate ($X_{14}$), demonstrating the robustness of the selection procedure. Additionally, the quality of the CATE estimates improves in Settings 2 and 3 compared to Setting 1, as reflected in both MSE and the rank correlation coefficient, highlighting the benefits of increased sample size and targeted variable inclusion.

}
\bibliographystyle{unsrtnat}
\bibliography{bibl.bib}

@article{vanderweele2014tutorial,
  title={A tutorial on interaction},
  author={VanderWeele, Tyler J and Knol, Mirjam J},
  journal={Epidemiologic Methods},
  volume={3},
  number={1},
  pages={33--72},
  year={2014},
  publisher={De Gruyter}
}

@misc{chen2025comparingmethodsassesstreatment,
      title={Comparing methods to assess treatment effect heterogeneity in general parametric regression models}, 
      author={Yao Chen and Sophie Sun and Konstantinos Sechidis and Cong Zhang and Torsten Hothorn and Björn Bornkamp},
      year={2025},
      eprint={2503.22548},
      archivePrefix={arXiv},
      primaryClass={stat.AP},
      url={https://arxiv.org/abs/2503.22548}, 
}

@article{harrel2001regression,
  title={Regression modeling strategies},
  author={Harrel, Frank E},
  journal={With Applications to Linear Models, Logistic Regression and Survival Analysis},
  pages={2001},
  year={2001}
}

@article{rube:shen:2015,
author = {Stephen J Ruberg and Lei Shen},
title = {Personalized Medicine: Four Perspectives of Tailored Medicine},
journal = {Statistics in Biopharmaceutical Research},
volume = {7},
pages = {214--229},
year = {2015},
}

@article{sechidis2025watchworkflowassesstreatment,
author = {Sechidis, Konstantinos and Sun, Sophie and Chen, Yao and Lu, Jiarui and Zhang, Cong and Baillie, Mark and Ohlssen, David and Vandemeulebroecke, Marc and Hemmings, Rob and Ruberg, Stephen and Bornkamp, Björn},
title = {WATCH: A Workflow to Assess Treatment Effect Heterogeneity in Drug Development for Clinical Trial Sponsors},
journal = {Pharmaceutical Statistics},
volume = {24},
number = {2},
pages = {e2463},
doi = {https://doi.org/10.1002/pst.2463},
url = {https://onlinelibrary.wiley.com/doi/abs/10.1002/pst.2463},
eprint = {https://onlinelibrary.wiley.com/doi/pdf/10.1002/pst.2463},
year = {2025}
}

@article{felson1993american,
  title={The American College of Rheumatology preliminary core set of disease activity measures for rheumatoid arthritis clinical trials},
  author={Felson, David T and Anderson, Jennifer J and Boers, Maarten and Bombardier, Claire and Chernoff, Miriam and Fried, Bruce and Furst, Daniel and Goldsmith, Charles and Kieszak, Stephanie and Lightfoot, Robert and others},
  journal={Arthritis \& Rheumatism: Official Journal of the American College of Rheumatology},
  volume={36},
  number={6},
  pages={729--740},
  year={1993},
  publisher={Wiley Online Library}
}

@article {Cardner2023,
	author = {Cardner, Mathias and Tuckwell, Danny and Kostikova, Anna and Forrer, Pascal and Siegel, Richard M and Marti, Alain and Vandemeulebroecke, Marc and Ferrero, Enrico},
	title = {Analysis of serum proteomics data identifies a quantitative association between beta-defensin 2 at baseline and clinical response to IL-17 blockade in psoriatic arthritis},
	volume = {9},
	number = {2},
	elocation-id = {e003042},
	year = {2023},
	doi = {10.1136/rmdopen-2023-003042},
	publisher = {BMJ Specialist Journals},
	URL = {https://rmdopen.bmj.com/content/9/2/e003042},
	eprint = {https://rmdopen.bmj.com/content/9/2/e003042.full.pdf},
	journal = {RMD Open}
}

@article{future1,
author = {Philip J. Mease  and Iain B. McInnes  and Bruce Kirkham  and Arthur Kavanaugh  and Proton Rahman  and Désirée van der Heijde  and Robert Landewé  and Peter Nash  and Luminita Pricop  and Jiacheng Yuan  and Hanno B. Richards  and Shephard Mpofu },
title = {Secukinumab Inhibition of Interleukin-17A in Patients with Psoriatic Arthritis},
journal = {New England Journal of Medicine},
volume = {373},
number = {14},
pages = {1329-1339},
year = {2015},
doi = {10.1056/NEJMoa1412679},

URL = {https://www.nejm.org/doi/full/10.1056/NEJMoa1412679},
}

@article{future2,
  title={{Secukinumab, a human anti-interleukin-{17A} monoclonal antibody, in patients with psoriatic arthritis {(FUTURE 2)}: a randomised, double-blind, placebo-controlled, phase 3 trial}},
  author={McInnes, Iain B and Mease, Philip J and Kirkham, Bruce and Kavanaugh, Arthur and Ritchlin, Christopher T and Rahman, Proton and Van der Heijde, Desiree and Landew{\'e}, Robert and Conaghan, Philip G and Gottlieb, Alice B and others},
  journal={The Lancet},
  volume={386},
  number={9999},
  pages={1137--1146},
  year={2015},
  publisher={Elsevier}
}

@article{future3,
  title={{Efficacy and safety of secukinumab administration by autoinjector in patients with psoriatic arthritis: results from a randomized, placebo-controlled trial (FUTURE 3)}},
  author={Nash, Peter and Mease, Philip J and McInnes, Iain B and Rahman, Proton and Ritchlin, Christopher T and Blanco, Ricardo and Dokoupilova, Eva and Andersson, Mats and Kajekar, Radhika and Mpofu, Shephard and others},
  journal={Arthritis research \& therapy},
  volume={20},
  number={1},
  pages={47},
  year={2018},
  publisher={Springer}
}

@article{future4,
  title={{Efficacy and safety of subcutaneous secukinumab 150 mg with or without loading regimen in psoriatic arthritis: results from the FUTURE 4 study}},
  author={Kivitz, Alan J and Nash, Peter and Tahir, Hasan and Everding, Andrea and Mann, He{\v{r}}man and Kaszuba, Andrzej and Pellet, Pascale and Widmer, Albert and Pricop, Luminita and Abrams, Ken},
  journal={Rheumatology and therapy},
  volume={6},
  number={3},
  pages={393--407},
  year={2019},
  publisher={Springer}
}

@article{future5,
  title={{Secukinumab improves active psoriatic arthritis symptoms and inhibits radiographic progression: primary results from the randomised, double-blind, phase III FUTURE 5 study}},
  author={Mease, Philip and van der Heijde, D{\'e}sir{\'e}e and Landew{\'e}, Robert and Mpofu, Shephard and Rahman, Proton and Tahir, Hasan and Singhal, Atul and Boettcher, Elke and Navarra, Sandra and Meiser, Karin and others},
  journal={Annals of the rheumatic diseases},
  volume={77},
  number={6},
  pages={890--897},
  year={2018},
  publisher={BMJ Publishing Group Ltd}
}

@article{lipkovich2017tutorial,
  title={Tutorial in biostatistics: data-driven subgroup identification and analysis in clinical trials},
  author={Lipkovich, Ilya and Dmitrienko, Alex and B D'Agostino Sr, Ralph},
  journal={Statistics in medicine},
  volume={36},
  number={1},
  pages={136--196},
  year={2017},
  publisher={Wiley Online Library}
}

@article{Lipkovich2024,
author = {Lipkovich, Ilya and Svensson, David and Ratitch, Bohdana and Dmitrienko, Alex},
title = {Modern approaches for evaluating treatment effect heterogeneity from clinical trials and observational data},
journal = {Statistics in Medicine},
volume = {43},
number = {22},
pages = {4388-4436},
doi = {https://doi.org/10.1002/sim.10167},
url = {https://onlinelibrary.wiley.com/doi/abs/10.1002/sim.10167},
eprint = {https://onlinelibrary.wiley.com/doi/pdf/10.1002/sim.10167},
year = {2024}
}

@Inbook{Wang2023,
author="Wang, Hongwei
and Feng, Dai
and Liu, Yingyi",
editor="He, Weili
and Fang, Yixin
and Wang, Hongwei",
title="Personalized Medicine with Advanced Analytics",
bookTitle="Real-World Evidence in Medical Product Development ",
year="2023",
publisher="Springer International Publishing",
address="Cham",
pages="289--320"
}

@article{Dahabreh2016,
    author = {Dahabreh, Issa J and Hayward, Rodney and Kent, David M},
    title = "{Using group data to treat individuals: understanding heterogeneous treatment effects in the age of precision medicine and patient-centred evidence}",
    journal = {International Journal of Epidemiology},
    volume = {45},
    number = {6},
    pages = {2184-2193},
    year = {2016},
    month = {11}
}

@article{Sanchez2022,
author = {Sanchez, Pedro  and Voisey, Jeremy P.  and Xia, Tian  and Watson, Hannah I.  and O’Neil, Alison Q.  and Tsaftaris, Sotirios A. },
title = {Causal machine learning for healthcare and precision medicine},
journal = {Royal Society Open Science},
volume = {9},
number = {8},
pages = {220638},
year = {2022},
doi = {10.1098/rsos.220638}
}

@article{bretz2014multiplicity,
  title={Multiplicity and replicability: two sides of the same coin},
  author={Bretz, Frank and Westfall, Peter H},
  journal={Pharmaceutical statistics},
  volume={6},
  number={13},
  pages={343--344},
  year={2014}
}

@article{Chung2013,
author = {EunYi Chung and Joseph P. Romano},
title = {{Exact and asymptotically robust permutation tests}},
volume = {41},
journal = {The Annals of Statistics},
number = {2},
publisher = {Institute of Mathematical Statistics},
pages = {484 -- 507},
year = {2013},
doi = {10.1214/13-AOS1090},
URL = {https://doi.org/10.1214/13-AOS1090}
}

@inproceedings{curth2021nonparametric,
  title={Nonparametric estimation of heterogeneous treatment effects: From theory to learning algorithms},
  author={Curth, Alicia and Van der Schaar, Mihaela},
  booktitle={International Conference on Artificial Intelligence and Statistics},
  pages={1810--1818},
  year={2021},
  organization={PMLR}
}

@article{jacob2021cate,
  title={Cate meets ml: Conditional average treatment effect and machine learning},
  author={Jacob, Daniel},
  journal={Digital Finance},
  volume={3},
  number={2},
  pages={99--148},
  year={2021},
  publisher={Springer}
}

@article{jacob2020cross,
  title={Cross-fitting and averaging for machine learning estimation of heterogeneous treatment effects},
  author={Jacob, Daniel},
  journal={arXiv preprint arXiv:2007.02852},
  year={2020}
}

@article{okasa2022meta,
  title={Meta-learners for estimation of causal effects: Finite sample cross-fit performance},
  author={Okasa, Gabriel},
  journal={arXiv preprint arXiv:2201.12692},
  year={2022}
}

@article{nowok2016synthpop,
 title={synthpop: Bespoke Creation of Synthetic Data in R},
 volume={74},
 url={https://www.jstatsoft.org/index.php/jss/article/view/v074i11},
 doi={10.18637/jss.v074.i11},
 number={11},
 journal={Journal of Statistical Software},
 author={Nowok, Beata and Raab, Gillian M. and Dibben, Chris},
 year={2016},
 pages={1–26}
}

@article{Curth2024,
author = {Curth, Alicia and Peck, Richard W. and McKinney, Eoin and Weatherall, James and van der Schaar, Mihaela},
title = {Using Machine Learning to Individualize Treatment Effect Estimation: Challenges and Opportunities},
journal = {Clinical Pharmacology \& Therapeutics},
volume = {115},
number = {4},
pages = {710-719},
doi = {https://doi.org/10.1002/cpt.3159},
year = {2024}
}

@article{kunzel2019metalearners,
  title={Metalearners for estimating heterogeneous treatment effects using machine learning},
  author={K{\"u}nzel, S{\"o}ren R and Sekhon, Jasjeet S and Bickel, Peter J and Yu, Bin},
  journal={Proceedings of the national academy of sciences},
  volume={116},
  number={10},
  pages={4156--4165},
  year={2019},
  publisher={National Acad Sciences}
}

@article{Neyman1923,
  title={On the application of probability theory to agricultural experiments. Essay on principles. Section 9. Roczniki Nauk Rolniczych TomX (in Polish) (1923)},
  author={Neyman, J.S.},
  journal={Translated in Statistical Science},
    volume={5},
  pages={465--480},
  year={1991},
  publisher={JSTOR}
}

@article{Rubin1974,
author = {{Rubin D. B.}},
journal = {Journal of Educational Psychology},
number = {5},
pages = {688--701},
title = {{Estimating causal effects of treatment in randomized and nonrandomized studies}},
volume = {66},
year = {1974}
}

@article{rubin2019essential,
  title={Essential concepts of causal inference: a remarkable history and an intriguing future},
  author={Rubin, Donald B},
  journal={Biostatistics \& Epidemiology},
  volume={3},
  number={1},
  pages={140--155},
  year={2019},
  publisher={Taylor \& Francis}
}

@article{imbens2004nonparametric,
  title={Nonparametric estimation of average treatment effects under exogeneity: A review},
  author={Imbens, Guido W},
  journal={Review of Economics and statistics},
  volume={86},
  number={1},
  pages={4--29},
  year={2004},
  publisher={MIT Press}
}

@article{Glynn_Quinn_2010, 
title={An Introduction to the Augmented Inverse Propensity Weighted Estimator}, 
volume={18}, 
DOI={10.1093/pan/mpp036}, 
number={1}, journal={Political Analysis}, 
author={Glynn, Adam N. and Quinn, Kevin M.}, 
year={2010}, 
pages={36–56}}

@article{hines2022demystifying,
  title={Demystifying statistical learning based on efficient influence functions},
  author={Hines, Oliver and Dukes, Oliver and Diaz-Ordaz, Karla and Vansteelandt, Stijn},
  journal={The American Statistician},
  volume={76},
  number={3},
  pages={292--304},
  year={2022},
  publisher={Taylor \& Francis}
}

@article{hines2022variable,
  title={Variable importance measures for heterogeneous causal effects},
  author={Hines, Oliver and Diaz-Ordaz, Karla and Vansteelandt, Stijn},
  journal={arXiv preprint arXiv:2204.06030},
  year={2022}
}

@book{gelman2020regression,
  title={Regression and other stories},
  author={Gelman, Andrew and Hill, Jennifer and Vehtari, Aki},
  year={2020},
  publisher={Cambridge University Press}
}

@manual{benchtm,
    title = {benchtm},
    author = {Sun, Sophie and Bornkamp, Bj{\"o}rn and Lu, Jiarui and Mirshani, Ardalan and Sechidis, Konstantinos and Chen, Yao},
    year = {2022},
    note = {R package},
    url = {https://github.com/Sophie-Sun/benchtm},
  }

@manual{grf,
    title = {grf: Generalized Random Forests},
    author = {Julie Tibshirani and Susan Athey and Erik Sverdrup and Stefan Wager},
    year = {2023},
    url = {https://CRAN.R-project.org/package=grf},
  }

@article{hoogland2021tutorial,
  title={A tutorial on individualized treatment effect prediction from randomized trials with a binary endpoint},
  author={Hoogland, Jeroen and IntHout, Joanna and Belias, Michail and Rovers, Maroeska M and Riley, Richard D and E. Harrell Jr, Frank and Moons, Karel GM and Debray, Thomas PA and Reitsma, Johannes B},
  journal={Statistics in medicine},
  volume={40},
  number={26},
  pages={5961--5981},
  year={2021},
  publisher={Wiley Online Library}
}

@article{sun2022comparing,
  title={Comparing algorithms for characterizing treatment effect heterogeneity in randomized trials},
  author={Sun, Sophie and Sechidis, Konstantinos and Chen, Yao and Lu, Jiarui and Ma, Chong and Mirshani, Ardalan and Ohlssen, David and Vandemeulebroecke, Marc and Bornkamp, Bj{\"o}rn},
  journal={Biometrical Journal},
  year={2022},
  publisher={Wiley Online Library}
}

@article{abecassis2024prediction,
  title={From prediction to prescription: Machine learning and Causal Inference},
  author={Ab{\'e}cassis, Judith and Dumas, Elise and Alberge, Julie and Varoquaux, Ga{\"e}l},
  year={2024}
}

@article{bornkamp2023predicting,
author = {Bornkamp, Björn and Zaoli, Silvia and Azzarito, Michela and Martin, Ruvie and Müller, Carsten Philipp and Moloney, Conor and Capestro, Giulia and Ohlssen, David and Baillie, Mark},
title = {Predicting subgroup treatment effects for a new study: Motivations, results and learnings from running a data challenge in a pharmaceutical corporation},
journal = {Pharmaceutical Statistics},
volume = {23},
number = {4},
pages = {495-510},
doi = {https://doi.org/10.1002/pst.2368},
url = {https://onlinelibrary.wiley.com/doi/abs/10.1002/pst.2368},
eprint = {https://onlinelibrary.wiley.com/doi/pdf/10.1002/pst.2368},
year = {2024}
}

@article{Lundberg2017,
author = {Scott Lundberg and Lee Su-In},
title = {A unified approach to interpreting model predictions.},
journal = {Advances in Neural Information Processing Systems},
volume = {30},
year = {2017}
}

@InProceedings{biasSHAP2023,
author="Baudeu, Raphael and Wright, Marvin N. and Loecher, Markus",
editor="Koprinska, Irena and Mignone, Paolo and Guidotti, Riccardo and Jaroszewicz, Szymon and Fr{\"o}ning, Holger and Gullo, Francesco and Ferreira, Pedro M. and Roqueiro, Damian and Ceddia, Gaia and Nowaczyk, Slawomir and Gama, Jo{\~a}o and Ribeiro, Rita and Gavald{\`a}, Ricard and Masciari, Elio and Ras, Zbigniew and Ritacco, Ettore and Naretto, Francesca and Theissler, Andreas and Biecek, Przemyslaw and Verbeke, Wouter and Schiele, Gregor and Pernkopf, Franz and Blott, Michaela and Bordino, Ilaria and Danesi, Ivan Luciano and Ponti, Giovanni and Severini, Lorenzo and Appice, Annalisa and Andresini, Giuseppina and Medeiros, Ib{\'e}ria and Gra{\c{c}}a, Guilherme and Cooper, Lee and Ghazaleh, Naghmeh and Richiardi, Jonas and Saldana, Diego and Sechidis, Konstantinos and Canakoglu, Arif and Pido, Sara and Pinoli, Pietro and Bifet, Albert and Pashami, Sepideh",
title="Are SHAP Values Biased Towards High-Entropy Features?",
booktitle="Machine Learning and Principles and Practice of Knowledge Discovery in Databases",
year="2023",
publisher="Springer Nature Switzerland",
address="Cham",
pages="418--433"
}

@article{seib:zeil:hoth:2016,
  author = {Heidi Seibold and Achim Zeileis and Torsten Hothorn},
  title = {Model-based Recursive Partitioning for Subgroup Analyses},
  journal = {International Journal of Biostatistics},
  year = 2016,
  doi = {10.1515/ijb-2015-0032},
  volume = 12,
  number = 1,
  pages = {45--63}
}

@article{athey2019estimating,
  title={Estimating treatment effects with causal forests: An application},
  author={Athey, Susan and Wager, Stefan},
  journal={arXiv preprint arXiv:1902.07409},
  year={2019}
}

@article{thomas:2018,
author = {Thomas, Marius and Bornkamp, Björn and Seibold, Heidi},
title = {Subgroup identification in dose-finding trials via model-based recursive partitioning},
journal = {Statistics in Medicine},
volume = {37},
number = {10},
pages = {1608-1624},
doi = {https://doi.org/10.1002/sim.7594},
url = {https://onlinelibrary.wiley.com/doi/abs/10.1002/sim.7594},
year = {2018}
}

@article{hooker2021unrestricted,
  title={Unrestricted permutation forces extrapolation: variable importance requires at least one more model, or there is no free variable importance},
  author={Hooker, Giles and Mentch, Lucas and Zhou, Siyu},
  journal={Statistics and Computing},
  volume={31},
  pages={1--16},
  year={2021},
  publisher={Springer}
}

@article{Boulesteix2011,
    author = {Boulesteix, Anne-Laure and Sauerbrei, Willi},
    title = {Added predictive value of high-throughput molecular data to clinical data and its validation},
    journal = {Briefings in Bioinformatics},
    volume = {12},
    number = {3},
    pages = {215-229},
    year = {2011},
    month = {01},
    doi = {10.1093/bib/bbq085}
}

@article{bo2023meta,
  title={A Meta-Learner Framework to Estimate Individualized Treatment Effects for Survival Outcomes},
  author={Bo, Na and Wei, Yue and Zeng, Lang and Kang, Chaeryon and Ding, Ying},
  year={2023}
}

@misc{li2024doublyrobusttargetedestimation,
      title={Doubly Robust Targeted Estimation of Conditional Average Treatment Effects for Time-to-event Outcomes with Competing Risks}, 
      author={Runjia Li and Victor B. Talisa and Chung-Chou H. Chang},
      year={2024},
      eprint={2407.18389},
      archivePrefix={arXiv},
      primaryClass={stat.ME},
      url={https://arxiv.org/abs/2407.18389}, 
}

@article{gao2024doubly,
  title={Doubly protected estimation for survival outcomes utilizing external controls for randomized clinical trials},
  author={Gao, Chenyin and Yang, Shu and Shan, Mingyang and Ye, Wenyu Wendy and Lipkovich, Ilya and Faries, Douglas},
  journal={arXiv preprint arXiv:2410.18409},
  year={2024}
}

@misc{ziersen2024variableimportancemeasuresheterogeneous,
      title={Variable importance measures for heterogeneous treatment effects with survival outcome}, 
      author={Simon Christoffer Ziersen and Torben Martinussen},
      year={2024},
      eprint={2412.11790},
      archivePrefix={arXiv},
      primaryClass={stat.ME},
      url={https://arxiv.org/abs/2412.11790}, 
}

@article{pryce2025causal,
  title={Causal machine learning for heterogeneous treatment effects in the presence of missing outcome data},
  author={Pryce, Matthew and Diaz-Ordaz, Karla and Keogh, Ruth H and Vansteelandt, Stijn},
  journal={Biometrics},
  volume={81},
  number={3},
  pages={ujaf098},
  year={2025},
  publisher={Oxford University Press}
}

@misc{hama2023modelfreevariableimportance,
      title={Model free variable importance for high dimensional data}, 
      author={Naofumi Hama and Masayoshi Mase and Art B. Owen},
      year={2023},
      eprint={2211.08414},
      archivePrefix={arXiv},
      primaryClass={cs.LG},
      url={https://arxiv.org/abs/2211.08414}, 
}

@article{verdinelli2024decorrelated,
  title={Decorrelated variable importance},
  author={Verdinelli, Isabella and Wasserman, Larry},
  journal={Journal of Machine Learning Research},
  volume={25},
  number={7},
  pages={1--27},
  year={2024}
}

@article{chernozhukov2017generic,
  title={Generic Machine Learning Inference on Heterogenous Treatment Effects in Randomized Experiments},
  author={Chernozhukov, Victor and Demirer, Mert and Duflo, Esther and Fern{\'a}ndez-Val, Iv{\'a}n},
  journal={arXiv preprint arXiv:1712.04802},
  year={2017}
}

@article{kennedy2023towards,
  title={Towards optimal doubly robust estimation of heterogeneous causal effects},
  author={Kennedy, Edward H},
  journal={Electronic Journal of Statistics},
  volume={17},
  number={2},
  pages={3008--3049},
  year={2023},
  publisher={The Institute of Mathematical Statistics and the Bernoulli Society}
}

@article{wager2018estimation,
  title={Estimation and inference of heterogeneous treatment effects using random forests},
  author={Wager, Stefan and Athey, Susan},
  journal={Journal of the American Statistical Association},
  volume={113},
  number={523},
  pages={1228--1242},
  year={2018},
  publisher={Taylor \& Francis}
}

@misc{Asher2024causalbounds,
      title={Model-Agnostic Covariate-Assisted Inference on Partially Identified Causal Effects}, 
      author={Wenlong Ji and Lihua Lei and Asher Spector},
      year={2024},
      eprint={2310.08115},
      archivePrefix={arXiv},
      primaryClass={econ.EM},
      url={https://arxiv.org/abs/2310.08115}, 
}

@book{ding2024first,
  title={A first course in causal inference},
  author={Ding, Peng},
  year={2024},
  publisher={CRC Press}
}

@article{fost:2011,
	title = {Subgroup identification from randomized clinical trial data},
	volume = {30},
	number = {24},
	journal = {Statistics in Medicine},
	author = {Foster, Jared C. and Taylor, Jeremy M.G. and Ruberg, Stephen J.},
	year = {2011},
	pages = {2867--2880}
}

@article{Floodgate2020,
  title={Floodgate: Inference for Model-Free Variable Importance},
  author={Zhang, Lu and Janson, Lucas},
  journal={arXiv preprint arXiv:2007.01283},
  year={2020}
}

@article{candes2018panning,
  title={Panning for gold:‘model-X’knockoffs for high dimensional controlled variable selection},
  author={Candes, Emmanuel and Fan, Yingying and Janson, Lucas and Lv, Jinchi},
  journal={Journal of the Royal Statistical Society Series B: Statistical Methodology},
  volume={80},
  number={3},
  pages={551--577},
  year={2018},
  publisher={Oxford University Press}
}

@article{liu2022fast,
  title={Fast and powerful conditional randomization testing via distillation},
  author={Liu, Molei and Katsevich, Eugene and Janson, Lucas and Ramdas, Aaditya},
  journal={Biometrika},
  volume={109},
  number={2},
  pages={277--293},
  year={2022},
  publisher={Oxford University Press}
}

@article{zimmermann2024all,
  title={All that Glitters Is not Gold: Type-I Error Controlled Variable Selection from Clinical Trial Data},
  author={Zimmermann, Manuela R and Baillie, Mark and Kormaksson, Matthias and Ohlssen, David and Sechidis, Konstantinos},
  journal={Clinical Pharmacology \& Therapeutics},
  volume={115},
  number={4},
  pages={774--785},
  year={2024},
  publisher={Wiley Online Library}
}

@misc{paillard2024,
      title={Measuring Variable Importance in Individual Treatment Effect Estimation with High Dimensional Data}, 
      author={Joseph Paillard and Vitaliy Kolodyazhniy and Bertrand Thirion and Denis A. Engemann},
      year={2024},
      eprint={2408.13002},
      archivePrefix={arXiv},
      primaryClass={cs.LG},
      url={https://arxiv.org/abs/2408.13002}, 
}

@MANUAL{R,
  title = {R: A Language and Environment for Statistical Computing},
  author = {{R Development Core Team}},
  organization = {R Foundation for Statistical Computing},
  address = {Vienna, Austria},
  year = {2011},
  note = {{ISBN} 3-900051-07-0}
}

@article{loh2019subgroup,
  title={Subgroup identification for precision medicine: A comparative review of 13 methods},
  author={Loh, Wei-Yin and Cao, Luxi and Zhou, Peigen},
  journal={Wiley Interdisciplinary Reviews: Data Mining and Knowledge Discovery},
  volume={9},
  number={5},
  pages={e1326},
  year={2019},
  publisher={Wiley Online Library}
}

@article{van2007super,
  title={Super learner},
  author={Van der Laan, Mark J and Polley, Eric C and Hubbard, Alan E},
  journal={Statistical applications in genetics and molecular biology},
  volume={6},
  number={1},
  year={2007},
  publisher={De Gruyter}
}

@Article{glmnet,
    title = {Regularization Paths for Generalized Linear Models via
      Coordinate Descent},
    author = {Jerome Friedman and Robert Tibshirani and Trevor Hastie},
    journal = {Journal of Statistical Software},
    year = {2010},
    volume = {33},
    number = {1},
    pages = {1--22},
    doi = {10.18637/jss.v033.i01},
  }

@article{hothorn2008implementing,
  title={Implementing a class of permutation tests: the coin package},
  author={Hothorn, Torsten and Hornik, Kurt and Van De Wiel, Mark A and Zeileis, Achim},
  journal={Journal of statistical software},
  volume={28},
  number={8},
  pages={1--23},
  year={2008},
  publisher={University of California at Los Angeles}
}

@article{hothorn2006lego,
  title={A lego system for conditional inference},
  author={Hothorn, Torsten and Hornik, Kurt and Van De Wiel, Mark A and Zeileis, Achim},
  journal={The American Statistician},
  volume={60},
  number={3},
  pages={257--263},
  year={2006},
  publisher={Taylor \& Francis}
}

@article{hothorn2006unbiased,
  title={Unbiased recursive partitioning: A conditional inference framework},
  author={Hothorn, Torsten and Hornik, Kurt and Zeileis, Achim},
  journal={Journal of Computational and Graphical statistics},
  volume={15},
  number={3},
  pages={651--674},
  year={2006},
  publisher={Taylor \& Francis}
}

@article{hothorn2015package,
  title={Package ‘party’},
  author={Hothorn, Torsten and Hornik, Kurt and Strobl, Carolin and Zeileis, Achim and Hothorn, Maintainer Torsten},
  journal={Package Reference Manual for Party Version 0.9-998},
  volume={16},
  pages={37},
  year={2015},
  publisher={Citeseer}
}

@article{strobl2007bias,
  title={Bias in random forest variable importance measures: Illustrations, sources and a solution},
  author={Strobl, Carolin and Boulesteix, Anne-Laure and Zeileis, Achim and Hothorn, Torsten},
  journal={BMC bioinformatics},
  volume={8},
  pages={1--21},
  year={2007},
  publisher={Springer}
}

@article{ImaiLi2024,
author = {Kosuke Imai and Michael Lingzhi Li},
title = {Statistical Inference for Heterogeneous Treatment Effects Discovered by Generic Machine Learning in Randomized Experiments},
journal = {Journal of Business \& Economic Statistics},
volume = {0},
number = {0},
pages = {1--13},
year = {2024},
publisher = {ASA Website},
doi = {10.1080/07350015.2024.2358909}
}

@article{debeer2020conditional,
  title={Conditional permutation importance revisited},
  author={Debeer, Dries and Strobl, Carolin},
  journal={BMC bioinformatics},
  volume={21},
  number={1},
  pages={1--30},
  year={2020},
  publisher={BioMed Central}
}

@article{Sechidis2021,
author = {Sechidis, Konstantinos and Kormaksson, Matthias and Ohlssen, David},
title = {Using knockoffs for controlled predictive biomarker identification},
journal = {Statistics in Medicine},
volume = {40},
number = {25},
pages = {5453-5473},
doi = {https://doi.org/10.1002/sim.9134},
year = {2021}
}

@article{Nie2020,
    author = {Nie, X and Wager, S},
    title = {Quasi-oracle estimation of heterogeneous treatment effects},
    journal = {Biometrika},
    volume = {108},
    number = {2},
    pages = {299-319},
    year = {2020},
    month = {09},
    issn = {0006-3444},
    doi = {10.1093/biomet/asaa076},
    url = {https://doi.org/10.1093/biomet/asaa076},
    eprint = {https://academic.oup.com/biomet/article-pdf/108/2/299/37938939/asaa076.pdf},
}

@article{van2024combining,
  title={Combining T-learning and DR-learning: a framework for oracle-efficient estimation of causal contrasts},
  author={van der Laan, Lars and Carone, Marco and Luedtke, Alex},
  journal={arXiv preprint arXiv:2402.01972},
  year={2024}
}

@article{kennedy2024semiparametric,
  title={Semiparametric doubly robust targeted double machine learning: a review},
  author={Kennedy, Edward H},
  journal={Handbook of statistical methods for precision medicine},
  pages={207--236},
  year={2024},
  publisher={Chapman and Hall/CRC}
}
\end{document}